\documentclass{ws-rv9x6}

\usepackage{subfigure}   
\usepackage{ws-rv-thm}   
\usepackage{ws-rv-van}   
\makeindex

\usepackage{amsmath}
\usepackage{amsfonts}
\usepackage{amssymb}
\usepackage{mathrsfs}
\usepackage{graphicx}
\usepackage{dcolumn}
\usepackage{bm}
\usepackage{xspace}
\usepackage{color}
\usepackage{calrsfs}

\DeclareFontFamily{OT1}{pzc}{}
\DeclareFontShape{OT1}{pzc}{m}{it}{<-> s * [1.10] pzcmi7t}{}
\DeclareMathAlphabet{\mathpzc}{OT1}{pzc}{m}{it}

\newcommand{\sqrts}{\sqrt{s}}
\newcommand{\sqrtsnn}{\sqrt{s_{_{\rm \textsc{nn}}}}}

\newcommand{\pT}{{p_{\textsc{t}}}}

\newcommand\cO{{\cal O}}

\newcommand{\LumiInt}{\mathcal{L}_{\mbox{\rm \tiny{int}}}}

\newcommand{\pp}{{\rm{pp}}}
\newcommand{\pn}{{\rm{pn}}}
\newcommand{\ppbar}{{\rm{p$\bar{\rm p}$}}}
\newcommand{\pA}{{\rm{pA}}}
\newcommand{\pN}{{\rm{pN}}}
\newcommand{\pPb}{{\rm{pPb}}}
\newcommand{\AaAa}{{\rm{AA}}}
\newcommand{\PbPb}{{\rm{PbPb}}}
\newcommand{\TpA}{T_{_{\rm pA}}}
\newcommand{\TAA}{T_{\rm \textsc{aa}}}

\newcommand{\TpAsq}{T^{2}_{_{\rm pA}}}

\newcommand{\dtwor}{{d^2r}}
\newcommand{\dtwob}{{d^2b}}

\newcommand{\pythia}{{\sc pythia}}
\newcommand{\herwig}{{\sc herwig++}}
\newcommand{\mcfm}{{\sc mcfm}}
\newcommand{\vbfnlo}{{\sc vbfnlo}}

\newcommand{\sigmasps}{\sigma_{{\rm \textsc{sps}}}}

\newcommand{\sigSPS}{\sigma^{{\rm \textsc{sps}}}}
\newcommand{\sigDPS}{\sigma^{{\rm \textsc{dps}}}}
\newcommand{\sigTPS}{\sigma^{{\rm \textsc{tps}}}}

\newcommand{\sigmaeff}{\sigma_{\rm eff}}

\newcommand{\sigmaeffdps}{\sigma_{\rm eff,\textsc{dps}}}
\newcommand{\sigmaefftps}{\sigma_{\rm eff,\textsc{tps}}}
\newcommand{\sigmaeffnps}{\sigma_{\rm eff,\textsc{nps}}}
\newcommand{\sigmaeffdpspA}{\sigma_{\rm eff,\textsc{dps},pA}}
\newcommand{\sigmaefftpspA}{\sigma_{\rm eff,\textsc{tps},pA}}
\newcommand{\sigmaeffdpsAA}{\sigma_{\rm eff,\textsc{dps,aa}}}

\newcommand{\sigmaDPSone}{\sigma^{{\rm \textsc{dps,1}}}}
\newcommand{\sigmaDPStwo}{\sigma^{{\rm \textsc{dps,2}}}}
\newcommand{\sigmaDPSthree}{\sigma^{{\rm \textsc{dps,3}}}}

\newcommand{\sigmaTPSone}{\sigma^{{\rm \textsc{tps,1}}}}
\newcommand{\sigmaTPStwo}{\sigma^{{\rm \textsc{tps,2}}}}
\newcommand{\sigmaTPSthree}{\sigma^{{\rm \textsc{tps,3}}}}
\newcommand{\sigmaTPSnine}{\sigma^{{\rm \textsc{tps,9}}}}

\newcommand{\QQbar}{\textsc{q}\overline{\textsc{q}}}
\newcommand{\ccbar}{c\overline{c}}
\newcommand{\bbbar}{b\overline{b}}
\newcommand{\jpsi}{J/\psi}

\newcommand{\sigmaDPSjpsijpsi}{\sigma^{{\rm \textsc{dps}}}_{{\jpsi\jpsi}}}

\newcommand{\NDPS}{\rm N^{{\rm {\textsc{dps}}}}}

\newcommand*{\eg}{e.g.\@\xspace}
\newcommand*{\ie}{i.e.\@\xspace}
\newcommand*{\cm}{c.m.\@\xspace}
\newcommand*{\aka}{aka.\@\xspace}

\def\order#1{\mathcal{O}{(#1)}}
\def\ttt#1{\texttt{\small #1}}
\newcommand{\toppp}{\ttt{Top++}}

\begin{document}


\chapter[Double, triple, and $n$-parton scatterings in high-energy proton \& nuclear collisions]{Double, triple, and $n$-parton scatterings in high-energy proton and nuclear collisions}\label{dde_snigirev}
\vspace{-0.5cm}
\author[David d'Enterria and Alexander Snigirev]{David d'Enterria$^1$ and Alexander Snigirev$^2$}
\address{$^1$CERN, EP Department, 1211 Geneva, Switzerland\\
$^2$Skobeltsyn Institute of Nuclear Physics, Lomonosov Moscow State University, 119991, Moscow, Russia}

\begin{abstract}
The framework to compute the cross sections for the production of particles with high mass and/or large transverse momentum 
in double- (DPS), triple- (TPS), and in general $n$-parton scatterings, from the corresponding single-parton ($\sigmasps$) values 
in high-energy proton-proton, proton-nucleus, and nucleus-nucleus is reviewed. The basic parameter of the factorized 
$n$-parton scattering ansatz is an effective cross section $\sigmaeff$ encoding all unknowns about the underlying 
generalized $n$-parton distribution in the proton (nucleon). 
In its simplest and most economical form, the $\sigmaeff$ parameter can be derived from the transverse parton profile of the colliding protons and/or nucleus, 
using a Glauber approach. Numerical examples for the cross sections and yields expected for the concurrent DPS or TPS production 
of heavy-quarks, quarkonia, and/or gauge bosons in proton and nuclear collisions at LHC and Future Circular Collider (FCC)
energies are provided. The obtained cross sections are based on perturbative QCD predictions for $\sigmasps$ 
at next-to-leading-order (NLO) or next-to-NLO (NNLO) accuracy including, when needed, nuclear modifications of the corresponding 
parton densities.
\end{abstract}
\body

\tableofcontents

\section{Introduction}\label{sec:intro}

The extended nature of hadrons and their growing parton densities when probed at increasingly higher collision energies,
make it possible to simultaneously produce multiple particles with large transverse momentum and/or mass 
($\sqrt{\pT^2+m^2}\gtrsim$~2~GeV) in independent multiparton interactions (MPI)~\cite{Sjostrand:1987su} in 
proton-(anti)proton (pp, \ppbar)~\cite{Bartalini:2011jp,Abramowicz:2013iva,Bansal:2014paa,Astalos:2015ivw,Proceedings:2016tff}, 
as well as in proton-nucleus (pA)~\cite{Strikman:2001gz,DelFabbro:2003tj,Frankfurt:2004kn,Cattaruzza:2004qb,DelFabbro:2004md,Cattaruzza:2005nv,Treleani:2012zi,Blok:2012jr,dEnterria:2012jam,Calucci:2013pza,dEnterria:2014lwk,dEnterria:2014mzh,dEnterria:2016yhy}, and nucleus (AA)~\cite{dEnterria:2013mrp,dEnterria:2014lwk,dEnterria:2014mzh} collisions.
Double-, triple-, and in general $n$-parton scatterings depend chiefly on the transverse overlap of the matter densities of the colliding hadrons, 
and provide valuable information on (i) the badly-known tridimensional (3D) profile of the partons inside the nucleon, 
(ii) the unknown energy evolution of the parton density as a function of impact parameter ($b$), and 
(iii) the  role of multiparton  (space, momentum, flavour, colour,...) correlations in the hadronic wave functions.
A good understanding of $n$-parton scattering (NPS) is not only useful to improve our knowledge of the 3D parton 
structure of the proton, but it is also of relevance for a realistic characterization of backgrounds in searches 
of new physics in rare final-states with multiple heavy-particles.

The interest in MPI has increased in the last years, not only as a primary source of particle production at 
hadron colliders~\cite{Sjostrand:2017cdm}, but also due to their role~\cite{dEnterria:2010xip} in the 
``collective'' partonic behaviour observed in ``central'' pp collisions, bearing close similarities to that 
measured in heavy-ion collisions~\cite{Khachatryan:2010gv,Aad:2015gqa}. As a matter of fact, 
the larger transverse parton density in a nucleus (with $A$ nucleons) compared to that of a proton, 
significantly enhances double (DPS) and triple (TPS) parton scattering 
cross sections coming from interactions where the colliding partons belong to the same or to different nucleons
of the nucleus (nuclei), providing thereby additional information on the underlying multiparton dynamics. 


Many final-states involving the concurrent production of \eg\ heavy-quarks ($c,b$), quarkonia ($\jpsi$, $\Upsilon$), 
jets, and gauge bosons ($\gamma$, W, Z) have been measured and found consistent with DPS at the Tevatron 
(see \eg\ early results from CDF~\cite{Abe:1997xk,Abe:1997bp} and more recent ones from D0~\cite{Abazov:2014fha,Abazov:2015nnn}) 
as well as at the LHC (see \eg\ the latest results from ATLAS~\cite{Aaboud:2016dea,Aaboud:2016fzt}, 
CMS~\cite{Chatrchyan:2013xxa,Khachatryan:2015pea} and LHCb~\cite{Aaij:2015wpa}).
The TPS processes, although not observed so far, have visible cross sections for charm and bottom in 
pp~\cite{dEnterria:2016ids} and pA~\cite{dEnterria:2016yhy} collisions at LHC and future circular 
(FCC)~\cite{Mangano:2016jyj,Dainese:2016gch} energies. The present writeup reviews and extends our past work on DPS and TPS in high-energy
pp, pA and AA collisions~\cite{dEnterria:2012jam,dEnterria:2013mrp,dEnterria:2014lwk,dEnterria:2014mzh,dEnterria:2016ids,dEnterria:2016yhy},
expanding the basic factorized formalism to generic NPS processes, and presenting realistic cross section estimates 
for the double- and triple-parton production of heavy-quarks, quarkonia, and/or gauge bosons in proton and nuclear collisions 
at LHC and FCC.

\section{\mbox{$n$-parton scattering cross sections in hadron-hadron collisions}}
\label{sec:2}

In a generic hadronic 
collision, the inclusive cross section to produce $n$ hard particles in $n$ independent hard parton scatterings, 
$h h' \to a_1\ldots a_n$, can be written as a convolution of generalized $n$-parton distribution functions (PDF) 
and elementary partonic cross sections summed over all involved partons,
\begin{eqnarray} 
\label{eq:master}
\sigma^{\rm \textsc{nps}}_{hh' \to a_1\ldots a_n} & = & \left(\frac{\mathpzc{m}}{n!}\right)\,\sum \limits_{i_1,..,i_n,i'_1,..,i'_n} 
\int \Gamma^{i_1\ldots i_n}_{h}(x_1,..,x_n; {\bf b_1},..,{\bf b_n}; Q^2_1,..,Q^2_n)\nonumber \\
& &\times\,\hat{\sigma}_{a_1}^{i_1i'_1}(x_1,x_1',Q^2_1)\,\cdots\,\hat{\sigma}_{a_n}^{i_ni'_n}(x_n,x_n',Q^2_n)\\
& &\times\,\Gamma^{i'_1...i'_n}_{h'}(x'_1,\ldots, x'_n; {\bf b_1} - {\bf b},\ldots,{\bf b_n} - {\bf b}; Q^2_1 ,\ldots, Q^2_n)\nonumber\\
& &\times\, dx_1 \ldots dx_n\,dx_1' ,\ldots, dx_n'\,d^2b_1 ,\ldots,d^2b_n\,d^2b.\nonumber
\end{eqnarray}
Here, $\Gamma^{i_1...i_n}_{h}(x_1,\ldots,x_n; {\bf b_1},\ldots, {\bf b_n}; Q^2_1,\ldots, Q^2_n)$ are $n$-parton 
generalized distribution functions, depending on the momentum fractions $x_1,\ldots,x_n$, and energy scales $Q_1,\ldots, Q_n$, 
at transverse positions ${\bf b_1},\ldots, {\bf b_n}$ of the $i_1,\ldots,i_n$ partons, producing final-state particles 
$a_1,\ldots,a_n$ with subprocess cross sections $\hat{\sigma}_{a_1}^{i_1i'_1}, \ldots, \hat{\sigma}_{a_n}^{i_ni'_n}$. 
The combinatorial $(\mathpzc{m}/n!)$ prefactor takes into account the different cases of (indistinguishable or not) final states. 
For a set of identical particles (\ie\ when $a_1 =\ldots=a_n$) we have $\mathpzc{m}=1$, whereas $\mathpzc{m}=2,3,6,\ldots$ for 
final-states with an increasing number of different particles produced. In the particular cases of interest here, we have
\begin{itemize}
\item DPS: $\mathpzc{m}=1$ if $a_1=a_2$; and $\mathpzc{m}=2$ if $a_1\neq a_2$. 
\item TPS: $\mathpzc{m}=1$ if $a_1=a_2=a_3$; $\mathpzc{m}=3$ if $a_1=a_2$, or $a_1=a_3$, or $a_2=a_3$; and $\mathpzc{m}=6$ if $a_1 \neq a_2 \neq a_3$.
\end{itemize}
The~$n$-parton~distribution~function $\Gamma^{i_1...i_n}_{h}(x_1,..,x_n;{\bf b_1},..,{\bf b_n};Q^2_1,..,Q^2_n)$ 
theoretically encodes all the 3D parton structure information of the hadron of relevance to compute the NPS cross sections, 
including the density of partons in the transverse plane and any intrinsic partonic correlations in kinematical and/or 
quantum-numbers spaces. Since $\Gamma^{i_1\ldots\,i_n}_{h}$ is potentially a very complicated object, 
one often resorts to simplified alternatives to compute NPS cross sections based on simpler quantities. 
As a matter of fact, without any loss of generality, any $n$-parton cross section can be always expressed in a 
more economical and phenomenologically useful form in terms of single-parton scattering (SPS) inclusive cross sections,
theoretically calculable in perturbative quantum chromodynamics (pQCD) approaches through collinear factorization~\cite{collinear} 
as a function of ``standard'' (longitudinal) PDF, $D^i_h(x,Q^2)$, at a given order of accuracy in the QCD coupling expansion
(next-to-next-to-leading order, NNLO, being the current state-of-the-art for most calculations): 
\begin{eqnarray} 
\sigSPS_{hh' \to a} = \sum \limits_{i_1,i_2} \int D^{i_1}_h(x_1; Q^2_1) \,\hat{\sigma}^{i_1i_2}_{a}(x_1, x_1') \,D^{i_2}_{h'}(x_1'; Q^2_1) dx_1 dx_1'\,.
\label{eq:hardS}
\end{eqnarray}
More precisely, any $n$-parton cross section can be expressed as the $n$th-product of the corresponding SPS cross sections for the 
production of each single final-state particle, normalized by the ($n$th$-1$) power of an effective cross section,
\begin{equation} 
\label{eq:master2}
\sigma_{hh' \to a_1\ldots\,a_n}^{\rm \textsc{nps}} = \left(\frac{\mathpzc{m}}{n!}\right)\, \frac{\sigma_{hh' \to a_1}^{\rm \textsc{sps}} 
\cdots\, \sigma_{hh' \to a_n}^{\rm \textsc{sps}}}{\sigmaeffnps^{n-1}},
\end{equation} 
where $\sigmaeffnps$ encodes all the unknowns related to the underlying generalized PDF. 
Equation~(\ref{eq:master2}) encapsulates the intuitive result that the probability to produce $n$ particles 
in a given inelastic hadron-hadron collision 
should be proportional to the $n$-product of probabilities to independently produce each one of them, 
normalized by the $n$th$-1$ power of an effective cross section to guarantee the proper units of the final result 
(\ref{eq:master2}).\footnote{Indeed, in the simplest DPS case, the probability to produce particles $a, b$ in a pp collision is:
$P_{{\rm pp} \to ab} = P_{{\rm pp} \to a} \cdot P_{{\rm pp} \to b} = 
\frac{\sigma_{{\rm pp} \to a}}{\sigma^{\rm inel}_{\rm pp}} \cdot \frac{\sigma_{{\rm pp} \to b}}{\sigma^{\rm inel}_{\rm pp}}$, 
which implies: $\sigma_{{\rm pp} \to a,b} = \frac{\sigma_{{\rm pp} \to a} \cdot \sigma_{{\rm pp} \to b}}{\sigmaeff}$, with 
$\sigmaeff\approx\sigma^{\rm inel}_{\rm pp}$.
In reality, the measured value of $\sigmaeff\approx$~15~mb is a factor of 2--3 lower (\ie\ the DPS probability is 2--3 times
{\it larger}) than the naive $\sigmaeff\approx\sigma^{\rm inel}_{\rm pp}$ expectation for typical ``hard'' (minijet) inelastic pp
partonic cross sections $\sigma^{\rm inel}_{\rm pp}\approx$~30--50~mb. This is so because the independent-scattering assumption does not
hold as the probability to produce a second particle is higher in low-impact-parameter (large transverse overlap) pp events 
where a first partonic scattering has already taken place.}

The value of $\sigmaeffnps$ in Eq.~(\ref{eq:master2}) can be theoretically estimated making a few common approximations. 
First, the $n$-PDF are commonly assumed to be factorizable in terms of longitudinal and transverse components, \ie\
\begin{eqnarray} 
\label{eq:DxF}
& &\Gamma^{i_1..i_n}_{h}(x_1,\ldots,x_n;{\bf b_1},\ldots,{\bf b_n};Q^2_1,\ldots,Q^2_n)\nonumber\\
& &= D^{i_1..i_n}_h(x_1, \ldots, x_n; Q^2_1, \ldots, Q^2_n) \cdot f({\bf b_1}) \cdots\! f({\bf b_n}),
\end{eqnarray} 
where $f({\bf b_1})$ describes the transverse parton density of the hadron, often considered a universal function 
for all types of partons, from which the corresponding hadron-hadron overlap function can be derived:
\begin{eqnarray} 
\label{eq:f}
T({\bf b}) = \int f({\bf b_1}) f({\bf b_1 -b})d^2b_1 \,,
\end{eqnarray} 
with the fixed normalization $\int T({\bf b})d^2b = 1$. Making the further assumption that the longitudinal components
reduce to the product of independent single PDF, 
\begin{equation}
D^{i_1..i_n}_h(x_1, \ldots, x_n; Q^2_1, \ldots, Q^2_n) = D^{i_1}_h(x_1; Q^2_1) \cdots\! D^{i_n}_h(x_n; Q^2_n)\,,
\end{equation}
the effective NPS cross section bears a simple geometric interpretation in terms of powers of the
inverse of the integral of the hadron-hadron overlap function over all impact parameters, 
\begin{eqnarray}
\label{eq:sigmaeffnps}
\sigmaeffnps=\left\{\int d^2b \,T^n({\bf b})\right\}^{-1/(n-1)}\,.
\end{eqnarray} 

\section{Double and triple parton scattering cross sections in hadron-hadron collisions}
\label{sec:pp}

The generalized expression (\ref{eq:master}) for the case of double-parton-scattering cross sections in hadron-hadron collisions, 
$hh' \to a_1a_2$, reads
\begin{eqnarray} 
\sigDPS_{hh'\to a_1a_2} =  \left(\frac{\mathpzc{m}}{2}\right) & \sum \limits_{i,j,k,l} & \int \Gamma_{h}^{ij}(x_1,x_2; {\bf b_1},{\bf b_2}; Q^2_1, Q^2_2)\nonumber\\
& \times & \hat{\sigma}^{ik}_{a_1}(x_1, x_1',Q^2_1) \cdot \hat{\sigma}^{jl}_{a_2}(x_2, x_2',Q^2_2)\\
& \times & \Gamma_{h'}^{kl}(x_1', x_2'; {\bf b_1} - {\bf b},{\bf b_2} - {\bf b}; Q^2_1, Q^2_2)\nonumber\\
& \times & dx_1 dx_2 dx_1' dx_2' d^2b_1 d^2b_2 d^2b\,.\nonumber
\label{eq:masterDPS}
\end{eqnarray}
Applying the ``master'' equations~(\ref{eq:master2}) and (\ref{eq:sigmaeffnps}) for $n=2$, one can express this cross section 
as a double product of independent single inclusive cross sections
\begin{equation} 
\label{eq:}
\sigma_{hh' \to a_1a_2}^{\rm \textsc{dps}} =  \left(\frac{\mathpzc{m}}{2}\right)\, \frac{\sigma_{hh' \to a_1}^{\rm \textsc{sps}} \cdot
\sigma_{hh' \to a_2}^{\rm \textsc{sps}}}{\sigmaeffdps},
\end{equation} 
where the effective DPS cross section (\ref{eq:sigmaeffnps}) that normalizes the double SPS product is
\begin{eqnarray} 
\label{eq:sigmaeffDPS}
\sigmaeffdps=\left[ \int d^2b\,T^2({\bf b})\right]^{-1} \,.
\end{eqnarray}

Similarly, the generic expression (\ref{eq:master}) for the TPS cross section for the process $hh' \to a_1a_2a_3$ reads~\cite{Snigirev:2016uaq}
\begin{eqnarray} 
\label{eq:masterTPS}
\sigma^{\rm \textsc{tps}}_{hh' \to a_1a_2a_3}  & = &
\left(\frac{\mathpzc{m}}{3!}\right) \sum \limits_{i,j,k,l,m,n} \int\Gamma^{ijk}_{h}(x_1,x_2,x_3;{\bf b_1},{\bf b_2},{\bf b_3};Q^2_1,Q^2_2, Q^2_3) \nonumber \\
& &\times\hat{\sigma}_{a_1}^{il}(x_1, x_1',Q^2_1) \cdot \hat{\sigma}_{a_2}^{jm}(x_2, x_2',Q^2_2) \cdot \hat{\sigma}_{a_3}^{kn}(x_3, x_3',Q^2_3)\nonumber\\
& &\times\Gamma^{lmn}_{h'}(x_1', x_2', x_3'; {\bf b_1} - {\bf b},{\bf b_2} - {\bf b},{\bf b_3} - {\bf b}; Q^2_1, Q^2_2, Q^2_3)\nonumber\\
& &\times dx_1 dx_2 dx_3 dx_1' dx_2' dx_3' d^2b_1 d^2b_2 d^2b_3 d^2b.
\end{eqnarray}
which can be reduced to a triple product of independent single inclusive cross sections
\begin{equation} 
\sigma_{hh' \to a_1a_2a_3  }^{\rm \textsc{tps}} =  \left(\frac{\mathpzc{m}}{3!}\right)\, \frac{\sigma_{hh' \to a_1}^{\rm \textsc{sps}} \cdot
\sigma_{hh' \to a_2}^{\rm \textsc{sps}} \cdot \sigma_{hh' \to a_3}^{\rm \textsc{sps}}}{\sigmaefftps^2},
\label{eq:sigmaTPS}
\end{equation} 
normalized by the {\em square} of an effective TPS cross section (\ref{eq:sigmaeffnps}), that amounts to~\cite{dEnterria:2016ids} 
\begin{eqnarray} 
\sigmaefftps^2=\left[ \int d^2b \,T^3({\bf b})\right]^{-1}\,,
\label{eq:sigmaeffTPS}
\end{eqnarray}

One can estimate the values of the effective DPS~(\ref{eq:sigmaeffDPS}) and TPS~(\ref{eq:sigmaeffTPS}) cross sections
via Eq.~(\ref{eq:f}) for different transverse parton profiles of the colliding hadrons, such as those typically implemented in the
modern \pp\ Monte Carlo (MC) event generators \pythia\,8~\cite{Sjostrand:2007gs}, and \herwig~\cite{Seymour:2013qka}. 
In \pythia\,8, the \pp\ overlap function as a function of impact parameter is often parametrized in the form:
\begin{eqnarray} 
T({\bf b})= \frac{m}{2\pi r^2_p \,\Gamma (2/m)} \exp{[-(b/r_p)^m]} \,, 
\label{eq:e,T}
\end{eqnarray}
normalized to one, $\int T({\bf b})d^2b = 1$, where $r_p$ is the characteristic ``radius'' of the proton, 
$\Gamma$ is the gamma function, and the exponent $m$ depends on the MC ``tune'' obtained from fits to the 
measured underlying-event activity and various DPS cross sections in pp collisions~\cite{Khachatryan:2015pea}. 
It varies between a pure Gaussian ($m=2$) to a more peaked exponential-like ($m=0.7, 1$) distribution. 
From the corresponding integrals of the square and cube of $T({\bf b})$, we obtain:
\begin{eqnarray} 
\sigmaeffdps & = & \left(\int d^2b~ T^2({\bf b})\right)^{-1}= 2\pi r^2_p\,\frac{2^{2/m}\Gamma(2/m)}{m}\,,\;\;{\rm and} \label{eq:e,T2}
\end{eqnarray}
\begin{eqnarray} 
\sigmaefftps & = & \left(\int d^2b~ T^3({\bf b})\right)^{-1/2}= 2\pi r^2_p\,\frac{3^{1/m}\Gamma(2/m)}{m}.\label{eq:e,T3}
\end{eqnarray}
From Eq.~(\ref{eq:e,T2}), in order to reproduce the experimental $\sigmaeffdps \simeq 15\pm 5$~mb value extracted in 
multiple DPS measurements at Tevatron~\cite{Abe:1997xk,Abe:1997bp,Abazov:2014fha,Abazov:2015nnn} and 
LHC~\cite{Astalos:2015ivw,Aaboud:2016dea,Aaboud:2016fzt,Chatrchyan:2013xxa,Khachatryan:2015pea,Aaij:2015wpa}, 
the characteristic proton ``radius'' parameter amounts to $r_p \simeq 0.11 \pm 0.02, 0.24 \pm 0.04, 0.49 \pm 0.08$~fm for exponents 
$m= 0.7, 1, 2$ as defined in Eq.~(\ref{eq:e,T}). 
The values of $\sigmaeffdps$ and $\sigmaefftps$, Eqs.~(\ref{eq:e,T2})--(\ref{eq:e,T3}), are of course closely related: 
$\sigmaefftps =(3/4)^{1/m}\cdot\sigmaeffdps$. Such a relationship is independent of the exact numerical value of the proton 
``size'' $r_p$, but depends on the overall shape of its transverse profile characterized by the exponent $m$.
For typical \pythia-8 $m= 0.7,1,2$ exponents tuned from experimental data~\cite{Khachatryan:2015pea}, one obtains 
$\sigmaefftps~=~[0.66,0.75,0.87]\times\sigmaeffdps$ respectively. 

The \herwig\ event generator uses an alternative parametrization of the proton profile described by the dipole fit of the 
two-gluon form factor in the momentum representation~\cite{Blok:2010ge}
\begin{eqnarray} 
F_{2g}({\bf q})=1/(q^2/m^2_g+1)^2,
\label{eq:dip}
\end{eqnarray}
where the gluon mass $m_g$ parameter characterizes the transverse momentum $q$ distribution of the proton, 
and the transverse density is obtained from its Fourier-transform:
$f({\bf b})=\int e^{-i{\bf b}\cdot{\bf q}}F_{2g}({\bf q})\frac{d^2q}{(2\pi)^2}$.
The corresponding DPS (\ref{eq:sigmaeffDPS}) and TPS (\ref{eq:sigmaeffTPS}) effective cross sections 
read~\cite{Seymour:2013qka}:
\begin{equation} 
\label{eq:sigmaeffdip}
\sigmaeffdps= \left[\int  F_{2g}^4(q)\frac{d^2q}{(2\pi)^2}\right]^{-1}=\frac{28 \pi}{m^2_g} , 
\end{equation}
and~\cite{dEnterria:2016ids} 
\begin{eqnarray} 
\sigmaefftps & = &  \bigg[\int (2\pi)^2 \delta({\bf q_1}+{\bf q_2}+{\bf q_3}) F_{2g}({\bf q_1}) F_{2g}({\bf q_2}) F_{2g}({\bf q_3}) \nonumber \\
& & \times F_{2g}({\bf -q_1}) F_{2g}({\bf -q_2}) F_{2g}({\bf- q_3}) \frac{d^2q_1}{(2\pi)^2}\frac{d^2q_2}{(2\pi)^2}\frac{d^2q_3}{(2\pi)^2}\bigg]^{-1/2}.\nonumber
\end{eqnarray}
Numerically integrating the latter and combining it with (\ref{eq:sigmaeffdip}), we obtain $\sigmaefftps~=~0.83\times\sigmaeffdps$, 
which is quite close to the value derived for the Gaussian \pp\ overlap function in \pythia\,8.
In order to reproduce the experimentally measured $\sigmaeffdps \simeq 15 \pm 5$~mb values, 
the characteristic proton ``size'' for this parametrization amounts to $r_g=1/m_g \simeq 0.13 \pm 0.02$~fm. 

Despite the wide range of proton transverse parton densities and associated effective radius parameters considered, 
we find that the $\sigmaefftps\lesssim\sigmaeffdps$ result is robust with respect to the underlying parton profile.
As a matter of fact, from the average and standard deviation of all typical parton transverse distributions studied in 
Ref.\cite{dEnterria:2016ids}, the following relationship between double and triple scattering effective 
cross sections can be derived:
\begin{equation}
\sigmaefftps = k\times\sigmaeffdps, \; {\rm with}\;\; k = 0.82\pm 0.11\,. 
\label{eq:TPS_DPS_factor}
\end{equation}
Thus, from the typical $\sigmaeffdps\simeq 15 \pm 5$ value extracted from a wide range of DPS measurements 
at Tevatron and LHC, the following numerical effective TPS cross section is finally obtained:
\begin{equation}
\sigmaefftps = 12.5 \pm 4.5 \;{\rm mb}.
\label{eq:TPS_factor}
\end{equation}


\subsection{\mbox{TPS cross sections in pp collisions: Numerical examples}}
\label{sec:TPS_pp_ex}

\begin{figure*}[hptb!]
\centering
\includegraphics[width=0.40\columnwidth,height=3.25cm]{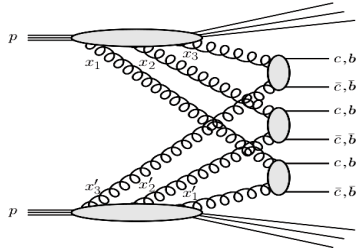}
\caption{Schematic diagram for the leading order contribution to triple charm ($\ccbar$) and bottom ($\bbbar$) 
pair production via gluon fusion, in TPS processes in \pp\ collisions.
\label{fig:TPS_HQ}}
\end{figure*}

Many theoretical and experimental studies exist that have extracted $\sigmaeffdps$ from computed and/or measured
DPS cross sections for a large variety of final-states in pp collisions~\cite{MPIbook}. In this subsection, 
we focuse therefore on the TPS case for which we presented the first-ever estimates in Ref~\cite{dEnterria:2016ids}. 
The experimental observation of triple parton scatterings in \pp\ collisions requires perturbatively-calculable 
processes with SPS cross sections not much smaller than $\mathcal{O}(\rm 1\;\mu b)$ since, otherwise, 
the corresponding TPS cross sections (which go as the cube of the SPS values) are extremely reduced. Indeed, 
according to Eq.~(\ref{eq:sigmaTPS}) with the data-driven estimate (\ref{eq:TPS_factor}), a triple hard process 
${\rm pp}\to a\,a\,a$, with SPS cross sections $\sigSPS_{{\rm pp}\to a}\approx 1\;\mu$b, has a very small 
cross section $\sigma^{\rm \textsc{tps}}_{{\rm pp} \to a\,a\,a}\approx 1$~fb. Evidence for TPS appears thereby 
challenging already without accounting for additional reducing factors arising from decay branching ratios,
and experimental acceptances and reconstruction inefficiencies, of the produced particles.
Promising processes to probe TPS, with not too small pQCD cross sections, are inclusive charm ($\pp\to\ccbar+X$), 
and bottom ($\pp\to\bbbar+X$), whose cross sections are dominated by gluon-gluon fusion ($gg\to\QQbar$) 
at small $x$, for which one can expect a non-negligible contributions of DPS~\cite{Luszczak:2011zp,Berezhnoy:2012xq,Cazaroto:2013fua} 
and TPS~\cite{dEnterria:2016ids,dEnterria:2016yhy,Maciula:2017meb} to their total inclusive production (Fig.~\ref{fig:TPS_HQ}). 

\begin{figure*}[htpb!]
\centering
\includegraphics[width=0.49\columnwidth]{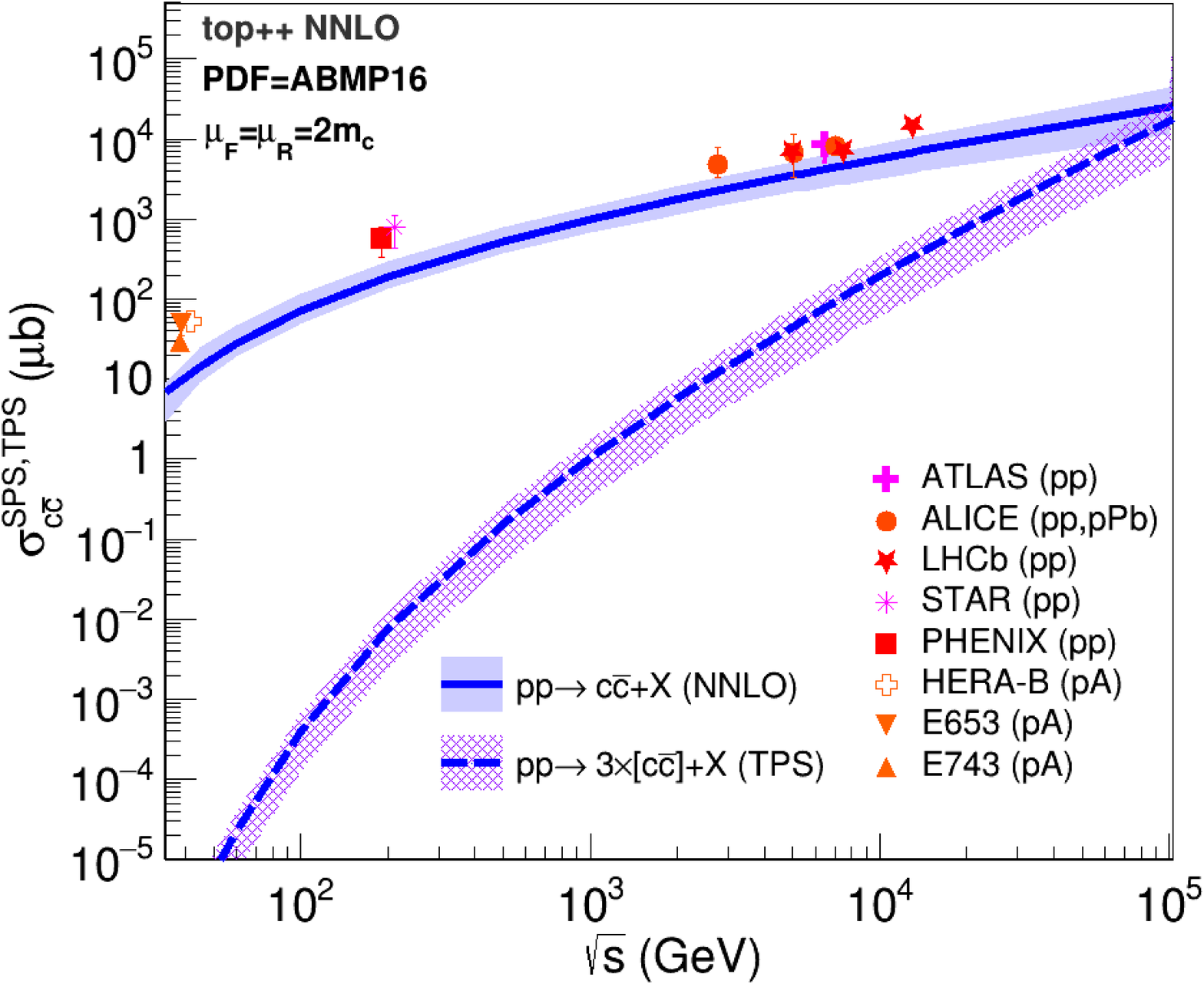}
\includegraphics[width=0.49\columnwidth]{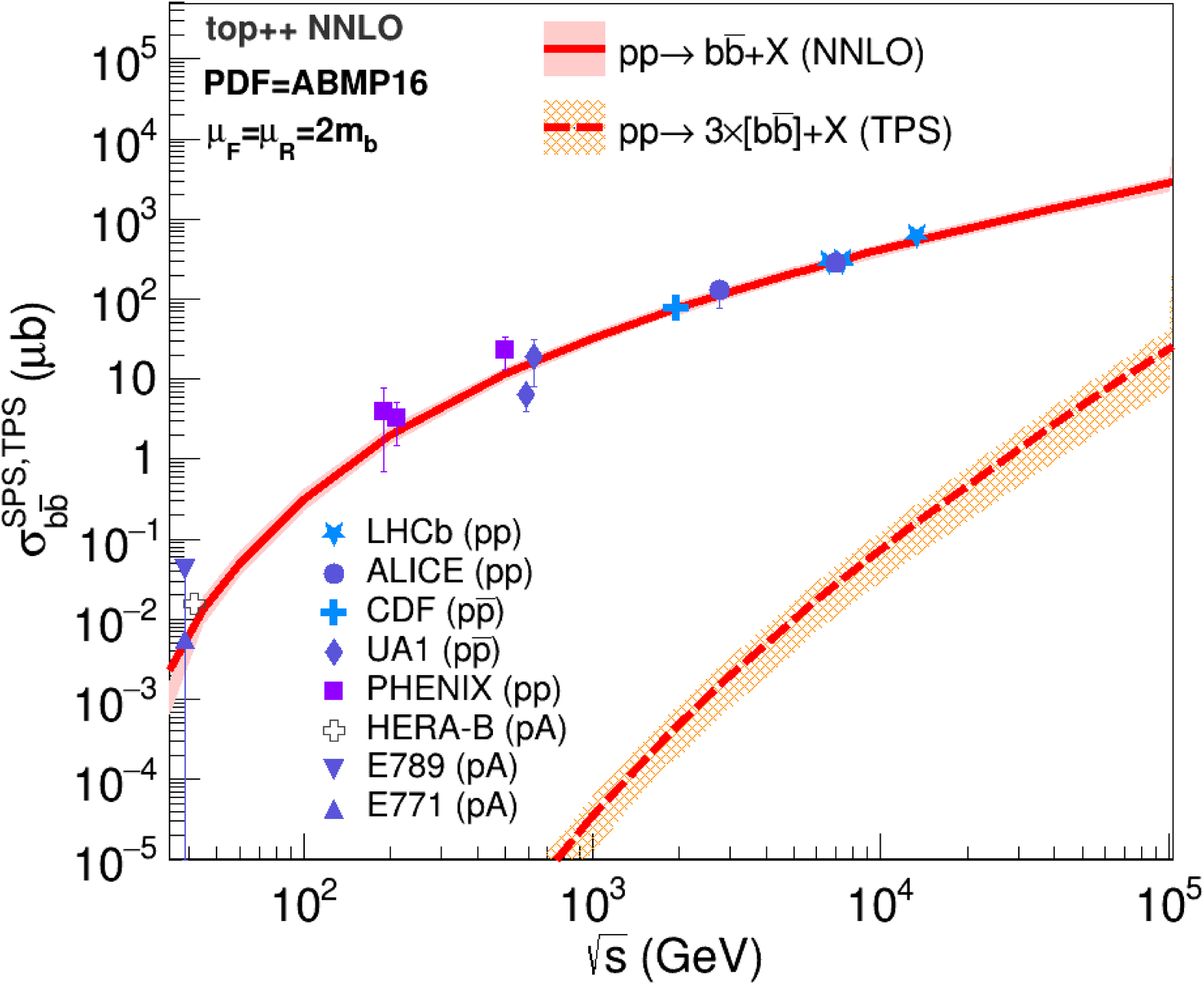}
\caption{Total charm (left) and bottom (right) cross sections in \pp\ collisions as a function of \cm\ energy, 
in single-parton (solid line) and triple-parton 
(dashed line) parton scatterings. Bands around curves indicate scale, PDF (and $\sigmaefftps$, in the case of $\sigTPS$) uncertainties added in quadrature.
The symbols are experimental data collected in~\cite{DdE}.
\label{fig:1}}
\end{figure*}

The TPS heavy-quark cross sections can be computed with Eq.~(\ref{eq:sigmaTPS}) for $\mathpzc{m}=1$, \ie\
$\sigma_{{\rm pp} \to \QQbar}^{\rm \textsc{tps}} = (\sigma_{{\rm pp} \to \QQbar}^{\rm \textsc{sps}})^3/(6\,\sigmaefftps^2)$
with $\sigmaefftps$ given by (\ref{eq:TPS_factor}), and $\sigma_{{\rm pp} \to \QQbar}^{\rm \textsc{sps}}$ 
calculated via Eq.~(\ref{eq:hardS}) at NNLO accuracy using a modified version~\cite{DdE} of the $\toppp$ (v2.0) 
code~\cite{Czakon:2013goa}, with $N_f=3,4$ light flavors, heavy-quark pole masses set at $m_{c,b}=1.67, 4.66$~GeV, 
default renormalization and factorization scales set at $\mu_{_{R}}=\mu_{_{F}}=2\, m_{c,b}$, and using the ABMP16 proton PDF~\cite{Alekhin:2016uxn}.
Such NNLO calculations increase the total SPS heavy-quark cross sections by up to 20\% at LHC energies
compared to the corresponding NLO results~\cite{fonll,mnr}, reaching a better agreement with the experimental data, 
and featuring much reduced scale uncertainties ($\pm50\%,\pm15\%$ for $\ccbar,\bbbar$)~\cite{DdE}.
Figure~\ref{fig:1} shows the resulting total SPS and TPS cross sections for charm and bottom production
over $\sqrts =$~35~GeV--100~TeV, and Table~\ref{tab:1} lists the results with associated uncertainties 
for the nominal pp \cm\ energies at LHC and FCC. The PDF uncertainties are obtained from the corresponding 28 eigenvalues 
of the ABMP16 set. The dominant uncertainty comes from the theoretical scales dependence, which is estimated 
by modifying $\mu_{_{R}}$ and $\mu_{_{F}}$ within a factor of two. 

\renewcommand\arraystretch{1.2}%
\begin{table}[htpb]
\tbl{\label{tab:1}Cross sections for charm and bottom production in SPS (NNLO) and TPS processes 
in \pp\ collisions at LHC and FCC energies. The quoted uncertainties include scales (sc), PDF, and total 
(quadratic, including $\sigmaefftps$) values.\vspace{0.15cm}}
{\begin{tabular}{@{}lccc@{}}\hline 
 Final state  &  $\sqrt{s}=14$ TeV & $\sqrt{s}=100$ TeV \\\hline
$\sigSPS_{\ccbar+X}$ & $7.1\pm3.5_{\rm sc}\pm0.3_{\rm \textsc{pdf}}$ mb &  $25.0\pm16.0_{\rm sc}\pm1.3_{\rm \textsc{pdf}}$ mb \\
$\sigTPS_{\ccbar\,\ccbar\,\ccbar+X}$ & $0.39\pm0.28_{\rm tot}$ mb &  $16.7\pm11.8_{\rm tot}$ mb \\\hline
$\sigSPS_{\bbbar+X}$ & $0.56\pm0.09_{\rm sc}\pm0.01_{\rm \textsc{pdf}}$ mb &  $2.8\pm0.6_{\rm sc}\pm0.1_{\rm \textsc{pdf}}$ mb \\
$\sigTPS_{\bbbar\,\bbbar\,\bbbar+X}$ & $0.19\pm0.12_{\rm tot}$ $\mu$b & $24\pm17_{\rm tot}$ $\mu$b\\
\hline 
\end{tabular}}
\end{table}

Figure~\ref{fig:1} shows that the TPS cross sections rise fast with $\sqrts$, 
as the cube of the corresponding SPS cross sections. Triple-$\ccbar$ production from three independent parton 
scatterings amounts to 5\% of the inclusive charm yields at the LHC ($\sqrts = 14$~TeV) and to more than half
of the total charm cross section at the FCC. Since the total \pp\ inelastic cross section at $\sqrts = 100$~TeV
is $\sigma_{\rm pp}\simeq$~105~mb~\cite{dEnterria:2016oxo}, charm-anticharm triplets are expected to be produced
in $\sim$15\% of the \pp\ collisions at these energies. Triple-$\bbbar$ cross sections remain quite small and reach only 
about 1\% of the inclusive bottom cross section at FCC(100~TeV). These results indicate that
TPS is experimentally observable in triple heavy-quark pair final-states at the LHC and FCC. The possibility of 
detecting triple charm-meson production in pp collisions at the LHC has been discussed in more detail in Ref.\cite{Maciula:2017meb}

\section{Double and triple parton scattering cross sections in proton-nucleus collisions}
\label{sec:pA}

In proton-nucleus collisions, the parton flux is enhanced by the number $A$ of nucleons in the nucleus and
the SPS cross section is simply expected to be that of proton-proton collisions or, more exactly, that of 
proton-nucleon collisions (\pN, with N\,=\,p,n being {\it bound} protons and neutrons with their appropriate relative 
fractions in the nucleus) taking into (anti)shadowing modifications of the nuclear PDF~\cite{Armesto:2006ph}, 
scaled by the factor $A$, \ie~\cite{d'Enterria:2003qs}
\begin{eqnarray} 
\sigSPS_{{\rm pA} \to a } = \sigSPS_{{\rm pN} \to a } \int \dtwob \, \TpA({\bf b}) = A \cdot \sigSPS_{{\rm pN} \to a }\,.
\label{eq:sigmaSPSpA}
\end{eqnarray} 
Here, $\TpA({\bf r})$ is the standard nuclear thickness function, analogous to Eq.~
(\ref{eq:f}) for the pp case, as a function of the impact parameter ${\bf r}$ between the colliding proton and nucleus, 
given by an integral of the nuclear density function $\rho_A({\bf r})$ over the longitudinal direction 
\begin{equation}
\TpA({\bf r}) = \int \rho_A\big(\sqrt{r^2+z^2}\big)\,dz, \;{\rm normalized \; to }\; \int \TpA({\bf r})\,\dtwor = A\,, 
\label{eq:TpA}
\end{equation}
which can be easily computed using (simplified) analytical nuclear profiles, and/or employing realistic 
Fermi-Dirac (\aka\ Woods-Saxon) nuclear spatial densities determined in elastic $e$A measurements~\cite{deJager}, 
via a MC Glauber model~\cite{d'Enterria:2003qs}.

The most naive assumption is to consider that the NPS cross sections in pA collisions can be obtained by simply $A$-scaling
the corresponding pp NPS values, as done via Eq.~(\ref{eq:sigmaSPSpA}) for the SPS cross sections. We show next that DPS 
and TPS cross sections in proton-nucleus collisions can be significantly enhanced, with extra $A^{4/3}$ (for DPS and TPS)
and $A^{5/3}$ (for TPS alone) terms complementing the $A$-scaling, due to additional multiple scattering probabilities among
partons from different nucleons.

\subsection{DPS cross sections in pA collisions}

\begin{figure}[hbtp!]
  \centering
  \includegraphics[width=0.90\columnwidth]{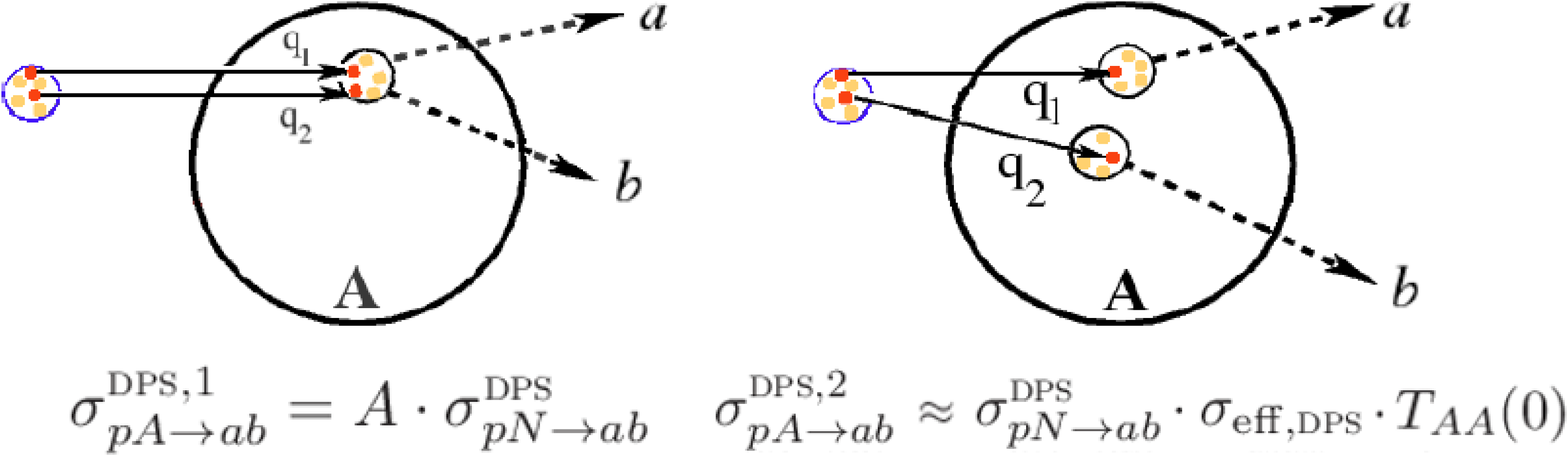}
  \caption{Schematic diagrams of double-parton scatterings contributions in pA collisions
   where the two colliding partons belong to the same (left) or a different (right) 
   pair of nucleons in the nucleus. The corresponding cross sections are described in the text.}
  \label{fig:diags_pA}
\end{figure}

The larger transverse parton density in nuclei compared to protons results in enhanced DPS cross sections, pA\,$\to ab$,
coming from interactions where the two partons of the nucleus belong to (1) the same nucleon, and (2) two different 
nucleons~\cite{Strikman:2001gz,DelFabbro:2003tj,Frankfurt:2004kn,Cattaruzza:2004qb,DelFabbro:2004md,Cattaruzza:2005nv,dEnterria:2012jam,Treleani:2012zi,Blok:2012jr}
as shown in Fig.~\ref{fig:diags_pA}. Namely,
\begin{equation} 
\sigDPS_{\rm pA} = \sigmaDPSone_{\rm pA} + \sigmaDPStwo_{\rm pA}\,,
\label{eq:sigmaDPS_pA}
\end{equation} 
where
\begin{enumerate}
\item The first term is just the $A$-scaled DPS cross section in \pN\ collisions: 
\begin{eqnarray} 
\sigmaDPSone_{{\rm pA} \to a b} = A \cdot \sigDPS_{{\rm pN} \to a b}\,,
\label{eq:sigmaDPS1}
\end{eqnarray} 
\item the second contribution, from parton interactions from two different nucleons, depends on the square of $\TpA$,
\begin{eqnarray} 
\label{eq:sigmaDPS2}
\sigmaDPStwo_{{\rm pA} \to a b} & = &\sigDPS_{{\rm pN} \to a b} \cdot \sigmaeffdps \cdot F_{\rm pA},\\
\mbox{ with } \; F_{\rm pA} & = & \frac{A-1}{A} \int \TpAsq({\bf r})\,\dtwor = (A-1)/A\cdot \TAA(0)\,,
\label{eq:FpA}
\end{eqnarray}
where the $(A-1)/A$ factor accounts for the difference between the number of nucleon pairs and the number of 
{\it different} nucleon pairs, and $\TAA(0)$ is the nuclear overlap function at b~=~0 for the corresponding AA collision. 
In the simplest approximation of a spherical nucleus with uniform nucleon density with radius $R_{\textsc{a}}\propto A^{1/3}$,
the factor (\ref{eq:FpA}) can be written as
\begin{eqnarray} 
F_{\rm pA} = \frac{9 A(A-1)}{8\,\pi\,R_{\textsc{a}}^2} \approx \frac{A^{4/3}}{14\,\pi}\;\,\rm{ [mb^{-1}]}\,.
\label{eq:FpAsimple}
\end{eqnarray} 
where the second approximate equality (valid for large $A$) indicates the corresponding dependence on the $A$ mass-number alone.
For Pb, with $A=208$ and $R_{\textsc{a}} \approx 7~\rm{fm} \approx 22~\rm{mb}^{1/2}$, one obtains $F_{\rm pA} \approx$~31.5~mb$^{-1}$,
in good agreement with the more accurate result, $F_{\rm pA} =$~30.25~mb$^{-1}$, computed with a Glauber MC~\cite{d'Enterria:2003qs} 
using the standard Woods-Saxon spatial density of the lead nucleus (radius $R_{\textsc{a}}$~=~6.36~fm, and surface thickness $a$~=~0.54~fm)~\cite{deJager}.
\end{enumerate}
The sum of (\ref{eq:sigmaDPS1}) and (\ref{eq:sigmaDPS2}) yields the inclusive cross section for the DPS production of
particles $a$ and $b$ in a \pA\ collision:
\begin{eqnarray} 
\sigDPS_{{\rm pA}\to a b} & = & A\cdot \sigDPS_{{\rm pN} \to a b}\left[1+\sigmaeffdps\,F_{\rm pA}/A  \right] \\
& \approx & A\cdot \sigDPS_{{\rm pN} \to a b}\left[1+\frac{\sigmaeffdps}{14\rm{\scriptstyle [mb]}\pi}A^{1/3}\right],
\label{eq:doublepA}
\end{eqnarray}
which is enhanced by the factor in parentheses compared to the $A$-scaled DPS cross section in
\pN\ collisions. Given the experimental $\sigmaeffdps\approx 15$~mb value, the pp-to-pA DPS 
enhancement factor can be further numerically simplified as $[1+A^{1/3}/\pi]$, which goes from $\sim$1.4 
for small to $\sim$3 for large nuclei. Namely, the relative weight of the two DPS terms of Eq.~(\ref{eq:doublepA}) 
goes from  $\sigma^{\rm \textsc{dps}, 1}_{{\rm pA} \to a b}:\sigma^{\rm \textsc{dps}, 2}_{{\rm pA} \to a b }
=0.7:0.3$ (small $A$) to $0.33:0.66$ (large $A$). Thus, \eg\ in the case of pPb collisions, 33\% of the DPS yields come 
from partonic interactions within just one nucleon of the Pb nucleus, whereas 66\% of them involve parton scatterings from 
two different Pb nucleons.

The final factorized DPS formula in proton-nucleus collisions can be written as a function of the 
elementary proton-nucleon single-parton cross sections as 
\begin{eqnarray} 
\sigDPS_{{\rm pA}\to a b} = \left(\frac{\mathpzc{m}}{2}\right) \frac{\sigSPS_{{\rm pN} \to a} \cdot \sigSPS_{{\rm pN} \to b}}{\sigmaeffdpspA}\,,
\label{eq:sigmapA_DPS}
\end{eqnarray}
where the effective DPS \pA\ cross section in the denominator depends on the effective cross section measured in pp, 
and on a pure geometric quantity ($F_{\rm pA}$) that is directly derivable from the well-known nuclear transverse profile, 
namely
\begin{eqnarray} 
\sigmaeffdpspA = \frac{\sigmaeffdps}{A+\sigmaeffdps\, F_{\rm pA}}  \approx \frac{\sigmaeffdps}{A+\sigmaeffdps\, \TAA(0)} 
\approx \frac{\sigmaeffdps}{A+A^{4/3}/\pi}. 
\label{eq:sigmaeffpA}
\end{eqnarray}
For a Pb nucleus (with $A$~=~208, and $F_{\rm pA} =$~30.25~mb$^{-1}$) and taking $\sigmaeffdps = 15 \pm 5$~mb, 
one obtains $\sigmaeffdpspA = 22.5 \pm 2.3$~$\mu$b. The overall increase of DPS cross sections in pA compared to pp collisions is
$\sigmaeffdps/\sigmaeffdpspA \approx [A+A^{4/3}/\pi]$ which, in the case of pPb implies a factor of 
$\sim$600 relative to pp (ignoring nuclear PDF effects here), \ie\ a factor of $[1+A^{1/3}/\pi]\approx$~3 
higher than the naive expectation assuming the same $A$-scaling of the single-parton cross sections, Eq.~(\ref{eq:sigmaSPSpA}). 
One can thus exploit such large expected DPS signals over the SPS backgrounds in proton-nucleus collisions to 
study double parton scatterings in detail and, in particular, to extract the value of $\sigmaeffdps$ independently 
of measurements in pp collisions---given that the parameter $F_{\rm pA}$ in Eq.~(\ref{eq:sigmaeffpA}) 
depends on the comparatively better-known transverse density of nuclei.

\subsubsection*{\mbox{DPS cross sections in pA collisions: Numerical examples}}

One of the ``cleanest'' channels to study DPS in pp collisions is same-sign WW production~\cite{Kulesza:1999zh} as it features 
precisely-known pQCD SPS cross sections, a clean experimental final-state with two like-sign leptons plus missing transverse
momentum from the undetected neutrinos, and small non-DPS backgrounds\footnote{The lowest order at which two same-sign 
W bosons can be produced is accompanied with two jets (W$^\pm$W$^\pm$jj), $q\,q \to $W$^\pm$W$^\pm \,q'\,q'$ with $q=u,c,\ldots$ and $q'=d,s,\ldots$
whose leading contributions are $\cO(\alpha_{\rm s}^2\alpha_{\rm w}^2)$ for the mixed QCD-electroweak
diagrams, and $\cO(\alpha_{\rm w}^4)$ for the pure vector-boson fusion (VBF) processes,
where $\alpha_{\rm w}$ is the electroweak coupling.}. The DPS cross section in pPb for same-sign WW production was
first estimated in Ref.\cite{dEnterria:2012jam}, computing the SPS W$^\pm$ cross sections ($\sigSPS_{{\rm pN} \to W}$) 
with \mcfm~(v.6.2)~\cite{mcfm,Campbell:2011bn} at NLO accuracy with CT10 proton~\cite{Lai:2010vv} 
and EPS09 nuclear~\cite{eps09} PDF, and setting default renormalisation and factorisation theoretical scales to $\mu = \mu_{_{R}} = \mu_{_{F}}$~=~$m_{_{W}}$.
The background W$^\pm$W$^\pm$jj cross sections are computed with \mcfm\ for the QCD part (formally at LO, but setting $\mu_{_{R}} = \mu_{_{F}}$~=~150~GeV 
to effectively account for missing higher-order corrections), and with \vbfnlo~(v.2.6)~\cite{vbfnlo,Arnold:2012xn} for the 
electroweak contributions with theoretical scales set to the momentum transfer of the exchanged boson, $\mu^2 = t_{_{W,Z}}$. 
In pPb at 8.8~TeV, the EPS09 nuclear PDF modifies the total W$^+$ (W$^-$) production cross section by about $-7\%~(+15\%)$ compared 
to that obtained using the free proton CT10 PDF~\cite{Paukkunen:2010qg}. 

We extend here the results of Ref.~\cite{dEnterria:2012jam}, using Eq.~(\ref{eq:sigmapA_DPS}) with $\mathpzc{m}$~=~1 and 
$\sigmaeffdpspA = 22.5 \pm 2.3$~$\mu$b, and including FCC pPb energies ($\sqrtsnn~=~63$~TeV).
The resulting cross sections are listed in Table~\ref{tab:2}.
The uncertainties of the SPS NLO single-W cross sections amount to about $\pm$10\% by adding in quadrature those from 
the EPS09 PDF eigenvector sets (the proton PDF uncertainties are much lower in the relevant $x,Q^2$ regions) and from
the theoretical scales (obtained by independently varying $\mu_{_{R}}$ and $\mu_{_{F}}$ within a factor of two). 
The QCD W$^\pm$W$^\pm$jj cross sections uncertainties are those from the full-NLO calculations~\cite{Melia:2010bm}, whereas 
those of the VBF cross sections are much smaller as they do not involve any gluons in the initial state. 
The DPS cross section uncertainties 
are dominated by a propagated $\pm$30\% uncertainty from $\sigmaeffdps$.

\begin{table}[htbp]
\tbl{\label{tab:2}Cross sections for the production of single-W, and same-sign W pairs 
in \pPb\ collisions at LHC and FCC \cm\ energies, computed at NLO with \mcfm\ and \vbfnlo\ for the processes quoted. 
The last column lists the same-sign DPS cross sections (sum of positive and negative W pairs) 
obtained with Eq.~(\ref{eq:sigmapA_DPS}) for $\sigmaeffdpspA = 22.5 \pm 2.3$~$\mu$b.
\vspace{0.15cm}}
{\begin{tabular}{lccc}\hline
 pPb  &  W$^+$, W$^-$  &  W$^+$W$^+$jj (QCD), (VBF)  &  W$^\pm$W$^\pm$ (DPS)  \\
  &  \small{NLO ($\mu$b)} \hspace{-0.8mm} &  \small{NLO (pb)}  &  \small{(pb)} \\\hline
 5.0~TeV  &  6.85 $\pm$ 0.68, 5.88 $\pm$ 0.59  & 12.1 $\pm$ 1.2, 12.4 $\pm$ 0.6  &  44 $\pm$ 13 \\
 8.8~TeV  &  12.6 $\pm$ 1.3, 11.1 $\pm$ 1.1    & 40.4 $\pm$ 4.0, 51.8 $\pm$ 2.0  &  152 $\pm$ 45 \\
 63~TeV   &  83.4 $\pm$ 8.4, 77.9 $\pm$ 7.8    & 166. $\pm$ 17., 2150. $\pm$ 220. &  6700. $\pm$ 2000. \\\hline
\end{tabular}}
\end{table}

\begin{figure}[hbtp!]
  \centering
  \includegraphics[width=0.65\columnwidth]{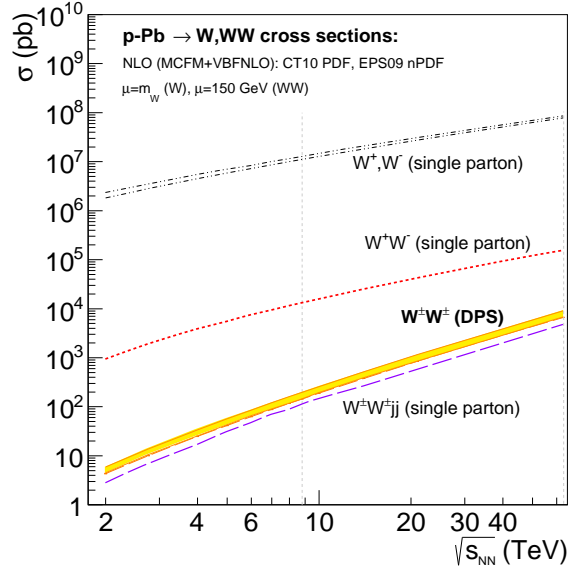} 
  \caption{Cross sections as a function of \cm\ energy for single-W, and W-pair (both opposite-sign and same-sign) 
   production from single-parton and from double-parton scatterings in pPb collisions. Dotted vertical lines
   indicate the nominal 8.8 and 63~TeV pPb energies at LHC and FCC.} 
  \label{fig:sigmaDPS_pPb_WW}
\end{figure}

Figure~\ref{fig:sigmaDPS_pPb_WW} shows the computed total cross sections for all W processes considered over the \cm\ 
energy $\sqrtsnn = 2$--65~TeV range. At the nominal LHC pPb \cm\ energy of 8.8~TeV, the same-sign WW DPS cross section 
is $\sigDPS_{\rm pPb\to WW}\approx$~150~pb (thick curve), 
larger than the sum of SPS backgrounds, 
$\sigSPS_{\rm pPb\to WWjj}$ (lowest dashed curve) obtained adding the QCD and electroweak cross sections for the production of
W$^+$W$^+$ (W$^-$W$^-$) plus 2 jets. In the fully-leptonic final-state (W$^\pm$W$^\pm\to\ell\nu\,\ell'\nu'$, with $\ell = \rm{e}^\pm, \mu^\pm$) 
and accounting for decay branching ratios and standard ATLAS/CMS acceptance and reconstruction cuts 
($|y^\ell|<2.5$, $\pT^\ell>15$~GeV), one expects up to 10 DPS same-sign WW events in $\cal{L}_{\rm int}$~=~2~pb$^{-1}$ 
integrated luminosity~\cite{dEnterria:2012jam}. At FCC energies ($\sqrtsnn = 63$~TeV), the ssWW DPS cross section is more than 
twice larger than the ssWW(jj) SPS one. With $\cal{L}_{\rm int} \approx$~30 pb$^{-1}$, and a factor twice larger rapidity 
coverage~\cite{Dainese:2016gch}, one expects $\order{10^4}$ ssWW pairs from DPS processes.
Same-sign WW production  in pPb collisions constitutes thereby a promising channel to measure $\sigmaeffdps$, 
independently of the standard pp-based extractions of this quantity.

\renewcommand{\arraystretch}{1.3}
\begin{table}[htbp!]
\tbl{\label{tab:3}Production cross sections at $\sqrtsnn = 8.8$~TeV for SPS quarkonia and electroweak bosons in pN collisions,
  and for DPS double-$\jpsi$, $\jpsi+\Upsilon$, $\jpsi$+W, $\jpsi$+Z,
  double-$\Upsilon$, $\Upsilon$+W, $\Upsilon$+Z, and same-sign WW, in pPb.
  DPS cross sections are obtained via Eq.~(\ref{eq:sigmapA_DPS}) for $\sigmaeffdpspA = 22.5$~$\mu$b
  (uncertainties, not quoted, are of the order of 30\%), and the corresponding yields, after dilepton decays 
   and acceptance+efficiency losses 
  (note that the $\jpsi$ yields are {\it per unit of rapidity} at mid- and forward-$y$, see text), 
  are given for the nominal 1~pb$^{-1}$ integrated luminosity.
  \vspace{0.25cm}}
{\begin{tabular}{lcccccccc}\hline
\pPb\ (8.8 TeV) & $\jpsi+\jpsi$ \hspace{0.5cm}& $\jpsi+\Upsilon$ \hspace{0.5cm}& $\jpsi$+W \hspace{0.5cm}& $\jpsi$+Z \hspace{0.5cm} \\\hline
$\sigSPS_{{\rm pN}\to a},\sigSPS_{{\rm pN}\to b}$  & 45~$\mu$b ($\times2$) & 45~$\mu$b, 2.6~$\mu$b & 45~$\mu$b, 60~nb & 45~$\mu$b, 35~nb \\
 $\sigDPS_{\rm pPb}$             & 45 $\mu$b &  5.2 $\mu$b & 120 nb & 70 nb \\
 $\NDPS_{\rm pPb}$ (1 pb$^{-1}$)\hspace{0.5cm}& $\sim$65 & $\sim$60 & $\sim$15 & $\sim$3 \\\hline
 & $\Upsilon+\Upsilon$ & $\Upsilon+$W & $\Upsilon+$Z & ss\,WW \\\hline
$\sigSPS_{{\rm pN}\to a},\sigSPS_{{\rm pN}\to b}$  & 2.6~$\mu$b ($\times2$) & 2.6~$\mu$b, 60~nb & 2.6~$\mu$b, 35~nb & 60~nb ($\times2$) \\
 $\sigDPS_{\rm pPb}$             & 150 nb & 7 nb & 4 nb & 150 pb \\
 $\NDPS_{\rm pPb}$ (1 pb$^{-1}$)&  $\sim$15 & $\sim$8 & $\sim$1.5 & $\sim$4 \\\hline
\end{tabular}}
\end{table}

Table~\ref{tab:3} collects the estimated DPS cross sections for the combined production of quarkonia ($\jpsi, \Upsilon$) 
and/or electroweak bosons (W,\,Z) in \pPb\ collisions at the nominal LHC energy of $\sqrtsnn = 8.8$~TeV. 
The quoted SPS pN cross sections have been obtained at NLO accuracy with the color evaporation model ({\sc cem})~\cite{Vogt:2012vr}
for quarkonia (see details in Section~\ref{sec:AA_DPS_ex}), and with \mcfm\ for the electroweak bosons, 
using CT10 proton and EPS09 nuclear PDF. 
The DPS cross sections are estimated via Eq.~(\ref{eq:sigmapA_DPS}) with $\sigmaeffdpspA = 22.5$~$\mu$b, and the visible
DPS yields are quoted  for $\cal{L}_{\rm int}$~=~1~pb$^{-1}$ integrated luminosities, 
taking into account the branching fractions BR($\jpsi,\Upsilon$,W,Z)~=~6\%, 2.5\%, 11\%, 3.4\% per dilepton decay; 
plus simplified acceptance and efficiency losses: ${\cal A\times E}(\jpsi$)~$\approx$~0.01 (over 1-unit of 
rapidity at $|y|=0$, and $|y|=2$),
and ${\cal A\times E}(\Upsilon;\rm W,Z)\approx$~0.2; 0.5 (over $|y|<2.5$). All listed processes 
are in principle observable in the LHC proton-lead runs, whereas rarer DPS processes like W+Z and Z+Z 
have much lower cross sections and require much higher luminosities and/or \cm\ energies such as those reachable at the FCC.

\subsection{TPS cross sections in pA collisions}
\label{sec:TPS_pA}

Similarly to the DPS case, the proton-nucleus TPS cross section for the pA\,$\to abc$ process, is obtained from the sum of three contributions:
\begin{equation} 
\sigTPS_{\rm pA} = \sigmaTPSone_{\rm pA} + \sigmaTPStwo_{\rm pA} + \sigmaTPSthree_{\rm pA}\,,\;{\rm with}
\label{eq:sigmaTPS_pA}
\end{equation} 
\begin{enumerate}
\item A cross section, scaling like Eq.~(\ref{eq:sigmaSPSpA}) for the SPS case, corresponding to the 
TPS value in pN collisions scaled by $A$, namely:
\begin{eqnarray} 
\sigma^{\rm \textsc{tps}, 1}_{{\rm pA} \to a b c} = A \cdot \sigma^{\rm \textsc{tps}}_{{\rm pN} \to a b c}\,.
\label{eq:doubleAB1}
\end{eqnarray} 
\item A second contribution, involving interactions of partons from two different nucleons in the
nucleus, depending on the square of $\TpA$,
\begin{eqnarray} 
\label{eq:doubleAB2}
\sigma^{\rm \textsc{tps}, 2}_{{\rm pA} \to a b c} = \sigma^{\rm \textsc{tps}}_{{\rm pN} \to a b c} \cdot 3 \, \frac{\sigmaefftps^2}{\sigmaeffdps} \, F_{\rm pA}, 
\label{eq:TpAsq}
\end{eqnarray} 
with $F_{\rm pA}$ given by Eq.~(\ref{eq:FpA}). 
\item A third term, involving interactions among partons from three different nucleons,
depending on the cube of $\TpA$,
\begin{eqnarray} 
\label{eq:abc3}
\sigma^{\rm \textsc{tps}, 3}_{{\rm pA} \to a b c} = \sigma^{\rm \textsc{tps}}_{{\rm pN} \to a b c} \cdot \sigmaefftps^2 \cdot C_{\rm pA}, \;\mbox{ with}\\
C_{\rm pA} = \frac{(A-1)(A-2)}{A^2}\int \dtwob\, \TpA^3({\bf b})\,, \label{eq:TpAcub}
\end{eqnarray} 
with the $(A-1)(A-2)/A^2$ factor introduced to take into account the difference between the 
total number of nucleon TPS and that of {\it different} nucleon TPS.
By using a hard-sphere approximation for a nucleus of radius $R_{\textsc{a}}\propto A^{1/3}$,
the $C_{\rm pA}$ factor can be analytically calculated as
\begin{equation}
C_{\rm pA}  = \frac{27}{4}\frac{A\,(A-1)\,(A-2)}{5 \pi^2 R_{\textsc{a}}^4} \approx \frac{A^{5/3}}{160\,\pi^2} \;\,\rm{ [mb^{-2}]}\,,
\end{equation} 
where the last approximate equality holds for large $A$. For a Pb nucleus ($A$~=~208, $R_{\textsc{a}} = 22$~mb$^{1/2}$) this
factor amounts to $C_{\rm pA}\approx$~5.1~mb$^{-2}$, in agreement with the $C_{\rm pA}$~=~4.75~mb$^{-2}$ numerically 
obtained through a Glauber MC with a realistic Woods-Saxon Pb profile.
\end{enumerate}
The inclusive TPS cross section for the independent production of three particles $a$, $b$, and $c$ in \pA\ collisions 
is obtained from the sum of the three terms (\ref{eq:doubleAB1}), (\ref{eq:doubleAB2}), and (\ref{eq:abc3}):
\begin{eqnarray} 
\label{eq:triplepA}
\sigTPS_{{\rm pA} \to a b c} & = & A \,\sigTPS_{\rm pN \to a b c} 
 \left[1+3 \, \frac{\sigmaefftps^2}{\sigmaeffdps} \frac{F_{\rm pA}}{A} + \sigmaefftps^2 \frac{C_{\rm pA}}{A} \right] \\
 & \approx & A\,\sigTPS_{\rm pN \to a b c}\left[1+\frac{\sigmaefftps^2}{\sigmaeffdps}\frac{3\,A^{1/3}}{14\rm{\scriptstyle [mb]}\pi}
 + \sigmaefftps^2 \frac{A^{2/3}}{160\rm{\scriptstyle [mb^2]}\pi^2} \right],
\end{eqnarray}
where the last approximation holds for large $A$, and can be written as a function of $\sigmaefftps$ and $A$ alone
making use of Eq.~(\ref{eq:TPS_DPS_factor}):
\begin{eqnarray} 
\label{eq:triplepA_bis}
\sigTPS_{{\rm pA} \to a b c} \approx A\,\sigTPS_{\rm pN \to a b c}\left[1+\sigmaefftps\frac{A^{1/3}}{5.7\rm{\scriptstyle [mb]}\pi}
 + \sigmaefftps^2 \frac{A^{2/3}}{160\rm{\scriptstyle [mb^2]}\pi^2} \right]\,.
\end{eqnarray}
The TPS cross section in pA collisions is enhanced by the factor in parentheses in Eqs.~(\ref{eq:triplepA})--(\ref{eq:triplepA_bis}) compared to the 
corresponding one in \pN\ collisions scaled by $A$. The final formula for TPS in proton-nucleus reads
\begin{equation} 
\sigma_{{\rm pA} \to abc}^{\rm \textsc{tps}} =   \left(\frac{\mathpzc{m}}{6}\right)\, \frac{\sigma_{{\rm pN} \to a}^{\rm \textsc{sps}} \cdot
\sigma_{{\rm pN} \to b}^{\rm \textsc{sps}} \cdot \sigma_{{\rm pN} \to c}^{\rm \textsc{sps}}}{\sigmaefftpspA^2}\,,
\label{eq:tripleabc}
\end{equation} 
where the effective TPS \pA\ cross section in the denominator depends on the effective TPS cross section measured in pp, 
and on purely geometric quantities ($F_{\rm pA}, C_{\rm pA}$) directly derivable from the well-known nuclear profiles~\cite{dEnterria:2016yhy},
\begin{eqnarray} 
\label{eq:sigmaeffpATPS}
\sigmaefftpspA &  = & \left[\frac{A}{\sigmaefftps^2} + \frac{3\,F_{\rm pA}[\rm{mb}^{-1}]}{\sigmaeffdps} + C_{\rm pA}[\rm{mb}^{-2}] \right]^{-1/2}
\end{eqnarray}
which can be numerically approximated  as a function of the number $A$ of nucleons in the nucleus (for $A$ large) alone, as follows
\begin{eqnarray} 
\label{eq:sigmaeffpATPSsphere}
\sigmaefftpspA & \approx & \left[\frac{A}{\sigmaefftps^2} + \frac{A^{4/3}}{5.7\rm{\scriptstyle[mb]}\,\pi\,\sigmaefftps} + \frac{A^{5/3}}{160\rm{\scriptstyle[mb^2]}\,\pi^2} \right]^{-1/2}\,.
\end{eqnarray} 
For a Pb nucleus ($A$~=~208, $F_{\rm pA} =$~30.25~mb$^{-1}$, and $C_{\rm pA}$~=~4.75~mb$^{-2}$) and taking $\sigmaefftps = 12.5 \pm 4.5$~mb, 
the effective TPS cross section amounts to $\sigmaefftpspA = 0.29 \pm 0.04$~mb.
Thus, for pPb the relative importance of the three TPS terms of Eq.~(\ref{eq:triplepA}) is 
$\sigma^{\rm \textsc{tps}, 1}_{{\rm pA} \to a b c}:\sigma^{\rm \textsc{tps}, 2}_{{\rm pA} \to a b c}:\sigma^{\rm \textsc{tps}, 3}_{{\rm pA} \to a b c}
=1:4.54:3.56$. Namely, in pPb collisions, 10\% of the TPS yields come from partonic interactions within just one nucleon 
of the lead nucleus, 50\% involve scatterings within two nucleons, and 40\% come from partonic interactions in three different
Pb nucleons. The sum of the three contributions in Eq.~(\ref{eq:triplepA}), ignoring differences between pN and pp collisions,
indicates that the TPS cross sections in pPb are about nine times larger than the naive expectation based on $A$-scaling 
of the corresponding pN TPS cross sections, Eq.~(\ref{eq:doubleAB1}). One can thus exploit the large expected TPS 
signals in proton-nucleus collisions to extract the $\sigmaefftps$ parameter, and thereby $\sigmaeffdps$ via Eq.~(\ref{eq:TPS_DPS_factor}), 
independently of TPS measurements in pp collisions---given that the $F_{\rm pA}$ and $C_{\rm pA}$ parameters in Eq.~(\ref{eq:triplepA}) 
depend on the comparatively better known transverse density of nuclei.


\subsubsection*{\mbox{TPS cross sections in pA collisions: Numerical examples}}

\begin{figure*}[htpb!]
\centering
\includegraphics[width=0.49\columnwidth]{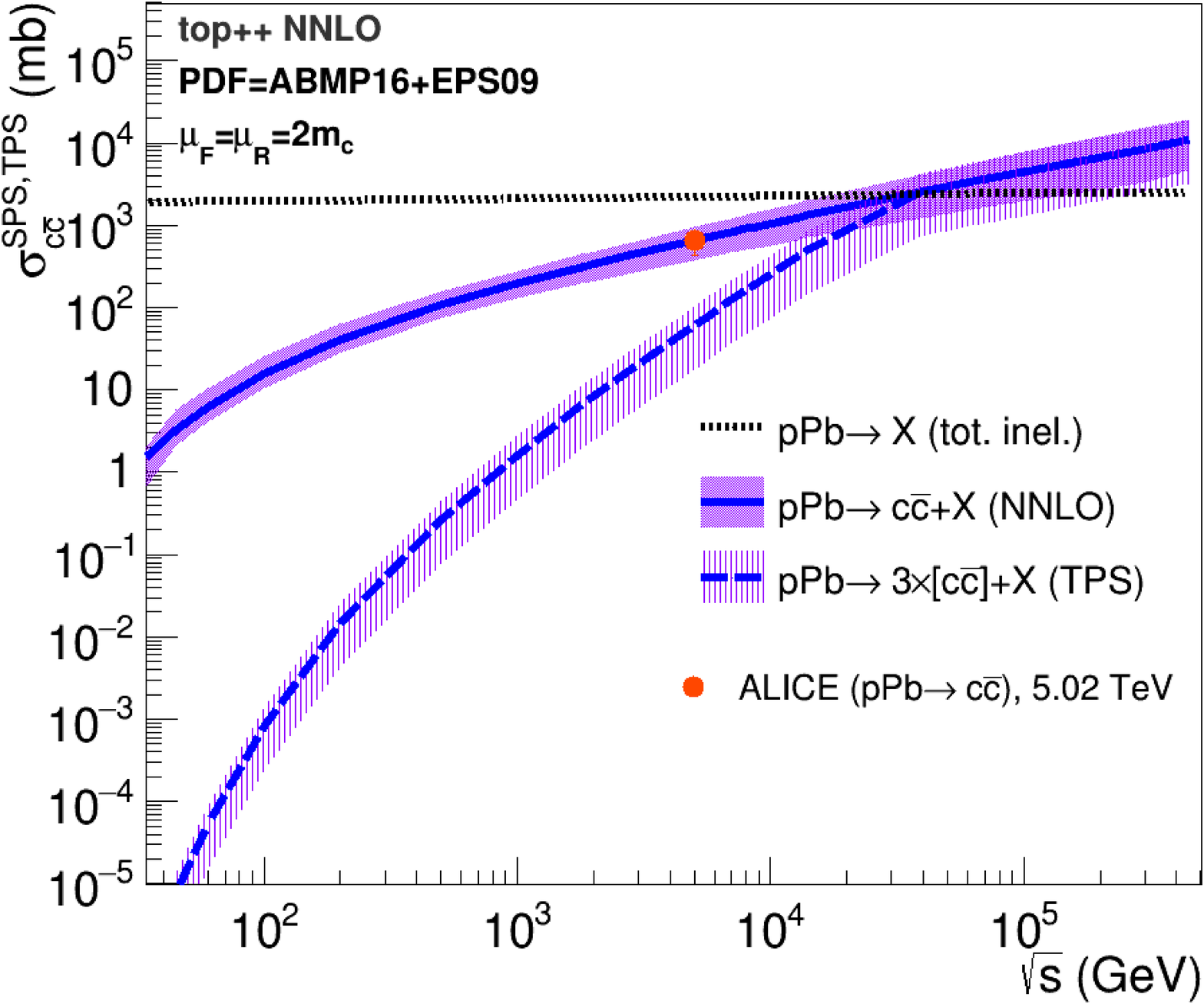}
\includegraphics[width=0.49\columnwidth]{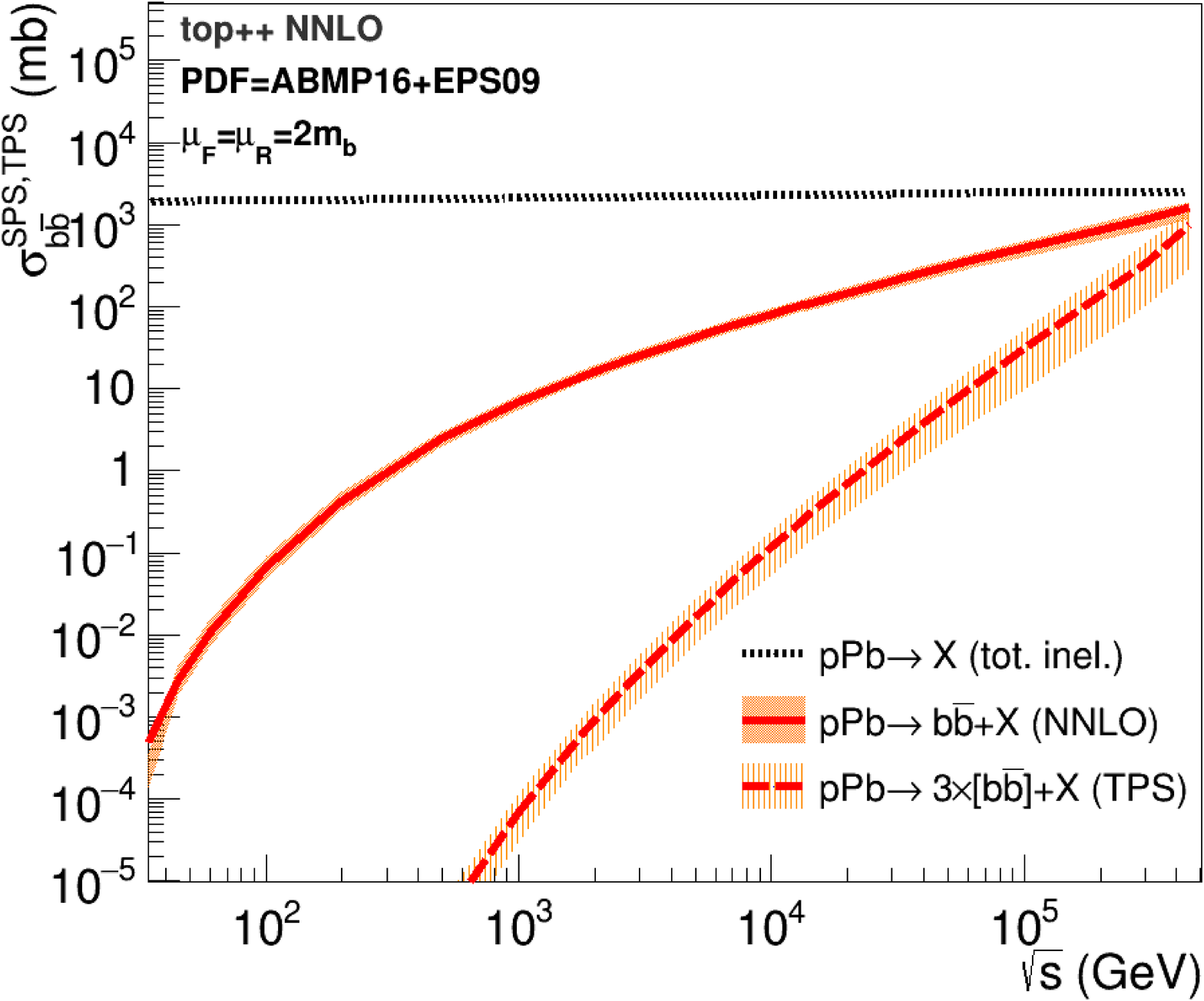}
\includegraphics[width=0.49\columnwidth]{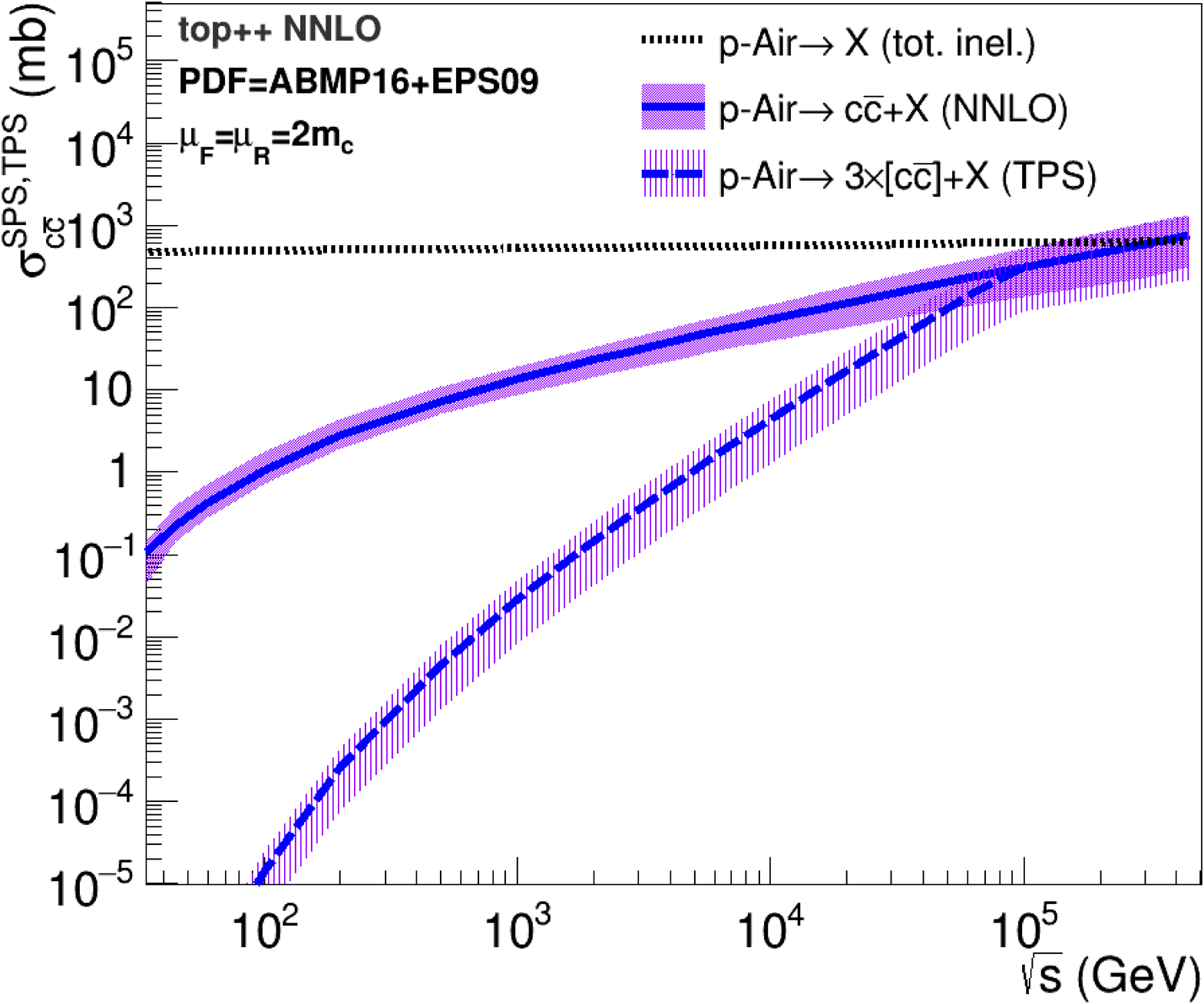}
\includegraphics[width=0.49\columnwidth]{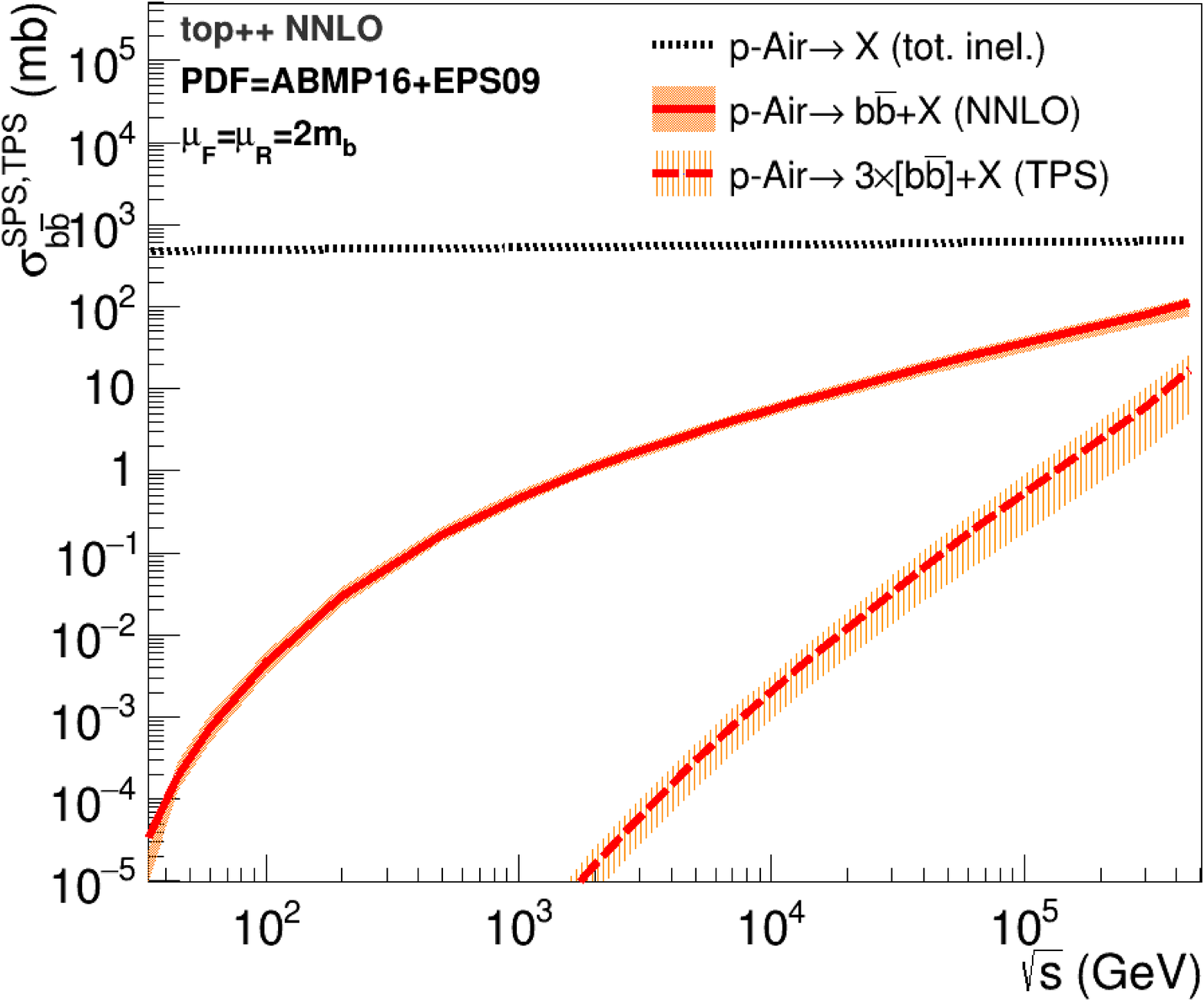}
\caption{Charm (left) and bottom (right) cross sections in pPb (top panels) and p-Air (bottom panels) collisions as a function of \cm\ energy, in single-parton 
(solid band) and triple-parton (dashed band) scatterings, compared to the total inelastic pA cross sections (dotted line in all panels). 
Bands around curves indicate scale, PDF (and $\sigmaefftps$, in the TPS case) uncertainties added in quadrature.
The ${\rm pPb}\to\ccbar+X$ charm data point on the top-left plot has been derived from the ALICE D-meson data~\cite{Adam:2016ich}.
\label{fig:TPS_pA}}
\end{figure*}

As a concrete numerical example in Ref.\cite{dEnterria:2016yhy} we have computed
the TPS cross sections for charm ($\ccbar$) and bottom ($\bbbar$) production,
following the motivation for the similar measurement in pp collisions (Section~\ref{sec:TPS_pp_ex}),
over a wide range of \cm\ energies, $\sqrtsnn \approx 5$--500~TeV, of 
relevance for collider (LHC and FCC) and ultra-high-energy cosmic rays physics. 
The TPS heavy-quark cross sections are computed via Eq.~(\ref{eq:tripleabc}) for $\mathpzc{m}=1$, \ie\ 
$\sigma_{{\rm pA} \to \ccbar,\bbbar}^{\rm \textsc{tps}} = (\sigma_{{\rm pN} \to \ccbar,\bbbar}^{\rm \textsc{sps}})^3/(6\,\sigmaefftpspA^2)$
with the effective TPS cross sections given by Eq.~(\ref{eq:sigmaeffpATPS}): $\sigmaefftpspA = 0.29 \pm 0.04$~mb for pPb, 
and $\sigmaefftpspA = 2.2 \pm 0.4$~mb for p-Air collisions\footnote{Using $A = 14.3$ for a 78\%--21\%  mixture of $^{14}$N--$^{16}$O, 
with $F_{{\rm pA}} = 0.51$~mb$^{-1}$, and $C_{{\rm pA}}~=~0.016$~mb$^{-2}$ obtained via a Glauber MC~\cite{d'Enterria:2003qs}.}. 
The SPS cross sections, $\sigma_{{\rm pN} \to \ccbar,\bbbar}^{\rm \textsc{sps}}$, are calculated at NNLO via Eq.~(\ref{eq:hardS}) 
with $\toppp$ (v.2.0) with the same setup as described in Sec.~\ref{sec:TPS_pp_ex},
using the ABMP16 proton and EPS09 nuclear PDF.
In the \pPb\ case, the inclusion of EPS09 nuclear shadowing reduces moderately the total charm and bottom 
cross sections in pN compared to pp collisions, by about 10\% (15\%) and 5\% (10\%) at the LHC (FCC). At $\sqrtsnn = 5.02$~TeV, our prediction 
($\sigma_{{\rm pPb}\to\ccbar}^{\rm \textsc{sps,nnlo}} = 650 \pm 290_{\rm sc} \pm 60_{\rm \textsc{pdf}}$~mb) agrees well with the ALICE total D-meson 
measurement~\cite{Adam:2016ich} extrapolated using FONLL~\cite{fonll} to a total charm cross section 
of $\sigma_{{\rm pPb}\to\ccbar}^{\rm \textsc{alice}} = 640 \pm 60_{\rm stat}\,^{+60}_{-110}\big|_{\rm syst}$~mb 
(data point in the top-left panel of Fig.~\ref{fig:TPS_pA}).
Since the TPS pPb cross section go as the cube of $\sigma_{{\rm pN}\to\QQbar}^{\rm \textsc{sps}}$, the impact of shadowing is 
amplified and leads to 15--35\% depletions of the TPS cross sections compared to results obtained with the free proton PDF.

\renewcommand\arraystretch{1.2}%
\begin{table}[htpb]
\tbl{\label{tab:4} Cross sections for inclusive inelastic, and for SPS and TPS charm and bottom production 
in pPb (at LHC and FCC energies) and p-Air (at GZK-cutoff \cm\ energies) collisions. For the SPS and TPS
cross sections the quoted values include scales, PDF, and total (quadratically added, including $\sigmaefftps$) 
uncertainties. [The asterisk indicates that the theoretical prediction of the TPS charm cross section is ``unphysical'' (see text).]
\vspace{0.15cm}}
{\begin{tabular}{lcccc} \hline
Process &  pPb(8.8 TeV) & pPb(63 TeV) & p-Air(430 TeV) \\ \hline
$\sigma^{\rm inel}_{\rm pA}$ & $2.2\pm0.4$ b &  $2.4\pm0.4$ mb &  $0.61\pm0.10$ b \\\hline
$\sigSPS_{\ccbar+X}$ & $0.96\pm0.45_{\rm sc}\pm0.10_{\rm \textsc{pdf}}$ b &  $3.4\pm1.9_{\rm sc}\pm0.4_{\rm \textsc{pdf}}$ b & $0.75\pm 0.5_{\rm sc}\pm 0.1_{\rm \textsc{pdf}}$ b\\
$\sigTPS_{\ccbar\,\ccbar\,\ccbar+X}$ & $200\pm140_{\rm tot}$ mb &  $8.7^*\pm6.2_{\rm tot}$ b &  $5.0^*\pm3.6_{\rm tot}$ b \\ \hline
$\sigSPS_{\bbbar+X}$ & $72\pm12_{\rm sc}\pm5_{\rm \textsc{pdf}}$ mb &  $370\pm75_{\rm sc}\pm30_{\rm \textsc{pdf}}$ mb &  $110\pm25_{\rm sc}\pm 5_{\rm \textsc{pdf}}$ mb \\
$\sigTPS_{\bbbar\,\bbbar\,\bbbar+X}$ & $0.084\pm 0.045_{\rm tot}$ $\mu$b & $11\pm7_{\rm tot}$ $\mu$b & $17 \pm 11_{\rm tot}$ $\mu$b \\ \hline
\end{tabular}}
\end{table}


Table~\ref{tab:4} collects the total inelastic and the heavy-quarks cross sections at $\sqrtsnn =$~8.8~TeV and 63~TeV in pPb collisions, 
and at $\sqrtsnn =$~430~TeV in p-Air collisions. The latter \cm\ energy corresponds to the so-called ``GZK cutoff''~\cite{Greisen:1966jv,Zatsepin:1966jv}
reached in collisions of ${\cal O}\rm (10^{20}\,eV)$ proton cosmic-rays, with N and O nuclei at rest in the upper atmosphere. 
The PDF uncertainties include those from the proton and nucleus in quadrature, as obtained from the corresponding 28$\oplus$30 
eigenvalues of the ABMP16$\oplus$EPS09 sets. The dominant uncertainty is linked to the theoretical scale choice, estimated 
by modifying $\mu_{_{R}}$ and $\mu_{_{F}}$ within a factor of two.
At the LHC, the large SPS $\ccbar$ cross section ($\sim$1~b) results in triple-$\ccbar$ cross sections from independent 
parton scatterings amounting to about 20\% of the inclusive charm yields. Since the total inelastic pPb cross sections 
is $\sigma^{\rm inel}_{\rm pPb}\approx$~2.2~b, charm TPS takes place in about 10\% of the pPb events at 8.8~TeV.
At the FCC, the theoretical TPS charm cross section even overcomes the inclusive charm one. Such an unphysical result indicates 
that quadruple, quintuple,...~parton-parton scatterings are expected to produce extra $\ccbar$ pairs with non-negligible probability.
The huge TPS $\ccbar$ cross sections in pPb at $\sqrtsnn = 63$~TeV, will make triple-$\jpsi$ production, with 
$\sigma(\jpsi\jpsi\jpsi+X)\approx$~1~mb, observable. Triple-$\bbbar$ cross sections remain comparatively small, in the 0.1~mb range, 
at the LHC but reach $\sim$10~mb (\ie\ 3\% of the total inclusive bottom cross section) at the FCC.

Figure~\ref{fig:TPS_pA} plots the cross sections over $\sqrtsnn \approx$~40~GeV--500~TeV for SPS (solid bands), TPS (dashed bands) 
for charm (left) and bottom (right) production, and total inelastic (dotted curve) in pPb (top panels) and p-Air (bottom panels) collisions. 
Whenever the central value of the theoretical TPS cross section overcomes the inclusive charm cross section, indicative of multiple 
(beyond three) $\ccbar$-pair production, we equalize it to the latter. At $\sqrtsnn \approx$~25~TeV, the total charm and 
inelastic pPb cross sections are equal implying that, above this \cm\ energy, {\it all} \pPb\ interactions produce at least three charm pairs. 
In the $\bbbar$ case, such a situation only occurs at much higher \cm\ energies, above 500~TeV. For p-Air collisions
at the GZK cutoff, the cross section for inclusive as well as TPS charm production equals the total inelastic cross section
($\sigma^{\rm inel}_{_{\rm pAir}}\approx$~0.61~b) indicating that {\it all} p-Air collisions produce at least
three $\ccbar$-pairs in multiple partonic interactions. In the $\bbbar$ case, about 20\% of the p-Air collisions produce bottom
hadrons, but only about 4\% of them have TPS production. These results emphasize the numerical importance of TPS processes in 
proton-nucleus collisions at colliders, and their relevance for hadronic MC models commonly used for the simulation of 
ultrarelativistic cosmic-ray interactions with the atmosphere~\cite{dEnterria:2011twh} which, so far, do not include any 
heavy-quark production.

\section{Double and triple parton scattering cross sections in nucleus-nucleus collisions}
\label{sec:AA}

In nucleus-nucleus collisions, the parton flux is enhanced by $A$ nucleons in each nucleus, and the SPS 
cross section is simply expected to be that of NN collisions, taking into account (anti)shadowing effects in the nuclear PDF, 
scaled by the factor $A^2$, \ie~\cite{d'Enterria:2003qs}
\begin{eqnarray} 
\sigSPS_{\textsc{aa} \to a} = \int \TAA({\bf b})d^2b = A^2 \cdot \sigSPS_{{\rm \textsc{nn}} \to a}\,.
\label{eq:sigmaSPSAA}
\end{eqnarray}
where $\TAA({\bf b})$ the standard nuclear overlap function, normalized to $A^2$,
\begin{equation}
\TAA({\bf b}) = \int \TpA({\bf b_1}) \TpA({\bf b_1-b}) d^2b_1 d^2b \;,
\end{equation}
with $\TpA({\bf b})$ being the nuclear thickness function at impact parameter ${\bf b}$, Eq.~(\ref{eq:TpA}), 
connecting the centres of the colliding nucleus in the transverse plane. In the next two subsections,
we present the estimates for DPS and TPS cross sections in AA collisions from the corresponding SPS values.

\subsection{\mbox{DPS cross sections in AA collisions}}
\label{sec:AA_DPS}

\begin{figure}[hbtp!]
  \centering
  \includegraphics[width=0.99\columnwidth]{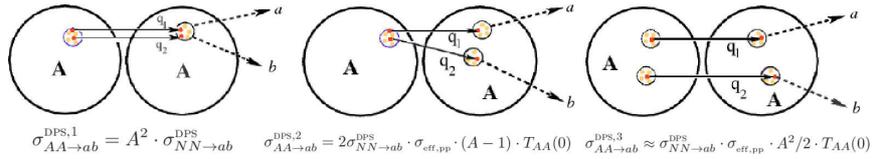} 
  \caption{Schematic diagrams contributing to DPS cross sections in AA collisions: 
   The two colliding partons belong to the same pair of nucleons (left), partons from one nucleon in 
   one nucleus collide with partons from two different nucleons in the other nucleus (center), and 
   the two colliding partons belong to two different nucleons from both nuclei (right).}
  \label{fig:diags_AA}
\end{figure}

The DPS cross section in AA is the sum of three terms, corresponding to the diagrams of Fig.~\ref{fig:diags_AA},
\begin{equation}
\sigDPS_{\textsc{aa}} = \sigmaDPSone_{\textsc{aa}} + \sigmaDPStwo_{\textsc{aa}} + \sigmaDPSthree_{\textsc{aa}}\,,{\rm where}
\end{equation}

\begin{enumerate}
\item The first term, similarly to the SPS cross sections Eq.~(\ref{eq:sigmaSPSAA}), is just the DPS cross section in NN collisions scaled by $A^2$,
\begin{eqnarray} 
\sigmaDPSone_{\textsc{aa} \to a b} = A^2 \cdot \sigDPS_{{\rm \textsc{nn}} \to a b}\,.
\label{eq:AAsigmaDPS1}
\end{eqnarray} 
\item The second term accounts for interactions of partons from one nucleon in one nucleus with partons 
from two different nucleons in the other nucleus,
\begin{eqnarray} 
\label{eq:AAsigmaDPS2}
\sigmaDPStwo_{\textsc{aa} \to a b} = 2\sigDPS_{{\rm \textsc{nn}} \to a b} \cdot \sigmaeffdps \cdot A\cdot F_{\rm pA}\,,
\end{eqnarray}
with $F_{\rm pA}\approx \TAA(0)$ given by Eq.~(\ref{eq:FpA}).
\item The third contribution from interactions of partons from two different nucleons in one nucleus with 
partons from two different nucleons in the other nucleus, reads
\begin{eqnarray} 
\label{eq:AAsigmaDPS3}
\sigmaDPSthree_{\textsc{aa} \to a b} = \sigDPS_{{\rm \textsc{nn}} \to a b} \cdot \sigmaeffdps \cdot T_{\rm 3,\textsc{aa}}\,\,{\rm with},
\end{eqnarray}
\begin{eqnarray}
\hspace{-2.cm}T_{\rm 3,\textsc{aa}} = & \left(\frac{A-1}{A}\right)^2 &\!\!\! \int\!\! \TpA({\bf b_1})\TpA({\bf b_2})\TpA({\bf b_1\!-\!b})\TpA({\bf b_2\!-\!b})d^2b_1 d^2b_2 d^2b \\ 
 = & \left(\frac{A-1}{A}\right)^2 & \int \dtwor\, \TAA^2({\bf r}) \approx \frac{A^2}{2}\cdot \TAA(0)\,,
\label{eq:AATpAsq3}
\end{eqnarray}
where the latter integral of the nuclear overlap function squared does not depend much on the precise shape of
the transverse parton density in the nucleus, amounting to $A^2/1.94\cdot \TAA(0)$ for a hard-sphere
and $A^2/2\cdot \TAA(0)$ for a Gaussian profile. The factor $((A-1)/A)^2$ takes into account 
the difference between the number of nucleon pairs and the number of {\it different} nucleon pairs.
\end{enumerate}
Adding (\ref{eq:AAsigmaDPS1}), (\ref{eq:AAsigmaDPS2}), and (\ref{eq:AAsigmaDPS3}), 
the inclusive cross section of a DPS process with two hard parton subprocesses $a$ and $b$ in \AaAa\
collisions 
can be written as  
\begin{eqnarray} 
\hspace{-0.5cm}\sigDPS_{\textsc{aa}\to a b}& = & A^2 \,\sigDPS_{{\rm \textsc{nn}} \to a b}\!\left[1\!+\!\frac{2}{A}\sigmaeffdps \,F_{\rm pA}+\frac{(A-1)^2}{A^2}\sigmaeffdps\!\! \int\!\! \dtwor\TAA^2({\bf r})\right] \label{eq:doubleAA1}\\
& \approx & A^2 \,\sigDPS_{{\rm \textsc{nn}} \to a b}\left[1+\frac{2}{A}\,\sigmaeffdps \,\TAA(0)\,+\,\frac{1}{2}\, \sigmaeffdps \, \TAA(0)\right] \\
& \approx & A^2 \,\sigDPS_{{\rm \textsc{nn}} \to a b}\left[1+ \frac{\sigmaeffdps}{7\rm{\scriptstyle [mb]}\,\pi}\,A^{1/3}\,+\,\frac{\sigmaeffdps}{28\rm{\scriptstyle [mb]}\,\pi}\,A^{4/3}\right] \,,
\label{eq:doubleAA}
\end{eqnarray}
where the last approximation, showing the $A$-dependence of the DPS cross sections, applies for large nuclei. 
The factor in parentheses in Eqs.~(\ref{eq:doubleAA1})--(\ref{eq:doubleAA})
indicates the enhancement in DPS cross sections in AA compared to the corresponding $A^2$-scaled 
values in nucleon-nucleon collisions, Eq.~(\ref{eq:AAsigmaDPS1}), which amounts to $\sim$27 (for small $A=40$)
or $\sim$215 (for large $A=208$).
The overall mass-number scaling of DPS cross sections in AA compared to pp collisions is given by
a $(A^2+k\,A^{7/3}+w\,A^{10/3})$ factor with $k,w\approx0.7,0.2$, which is clearly dominated numerically by the $A^{10/3}$ term. 
The final DPS cross section ``pocket formula'' in heavy-ion collisions 
can be written as 
\begin{eqnarray} 
\sigDPS_{\textsc{aa}\to ab} = \left(\frac{\mathpzc{m}}{2}\right) \frac{\sigSPS_{{\rm \textsc{nn}} \to a} \cdot \sigSPS_{{\rm \textsc{nn}} \to b}}{\sigmaeffdpsAA},
\label{eq:sigmaAADPS}
\end{eqnarray}
with the effective \AaAa\ normalization cross section amounting to
\begin{eqnarray} 
\sigmaeffdpsAA \approx \frac{1}{A^2\left[\sigmaeffdps^{-1}+\frac{2}{A}\,\rm{T}_{\textsc{aa}}(0)\,+\,\frac{1}{2}\,\TAA(0)\right]} \,.
\label{eq:sigmaeffAA}
\end{eqnarray}
For a value of $\sigmaeffdps \approx$~15~mb and for nuclei with mass numbers $A = 40$--240,
we find that the relative weights of the three components contributing to DPS scattering in AA collisions are 
$1:2.3:23$ (for $A = 40$) and $1:4:200$ (for $A = 208$). Namely, only 13\% (for $^{40}$Ca+$^{40}$Ca) or 
2.5\% (for $^{208}$Pb+$^{208}$Pb) of the DPS yields in AA collisions come from the first two diagrams of 
Fig.~\ref{fig:diags_AA} involving partons from one single nucleon. 
Clearly, the ``pure'' DPS contributions arising from partonic collisions within a single nucleon (first and second terms 
of Eq.~(\ref{eq:doubleAA})) are much smaller than the last term from double particle production coming from 
two independent {\it nucleon-nucleon} collisions. The DPS cross sections in \AaAa\ are practically unaffected by the value of 
$\sigmaeffdps$, but dominated instead by double-parton interactions from {\it different nucleons} in both nuclei. 
In the case of $^{208}$Pb\,-$^{208}$Pb collisions, the numerical value of Eq.~(\ref{eq:sigmaeffAA}) is 
$\sigmaeffdpsAA = 1.5 \pm 0.1$~nb, with uncertainties dominated by those of the Glauber MC determination of $\TAA(0)$.
Whereas the single-parton cross sections in \PbPb\ collisions, Eq.~(\ref{eq:sigmaSPSAA}), are enhanced by a 
factor of $A^2~\simeq~4\cdot 10^4$ compared to that in pp collisions, the corresponding double-parton cross 
sections are enhanced by a much higher factor of $\sigmaeffdps\,/\sigmaeffdpsAA \propto 0.2\,A^{10/3} \simeq 10^7$.


\subsubsection*{Centrality dependence of DPS cross sections in AA collisions}

The DPS cross sections discussed above are for ``minimum bias'' AA collisions without any selection in
reaction centrality. The cross sections for single and double-parton scattering within an impact-parameter interval
[b$_1$,b$_2$], corresponding to a given centrality percentile f$_{\%}$ of the total \AaAa\ cross section $\sigma^{\rm inel}_{\rm \textsc{aa}}$, 
with average nuclear overlap function $\langle \TAA[b_1,b_2]\rangle$ read (for large $A$, so that $A-1\approx A$):
\begin{eqnarray} 
\hspace{-0.5cm}\sigSPS_{\textsc{aa}[b_1,b_2] \to a} &=& A^2 \, \sigSPS_{{\rm \textsc{nn}} \to a}\, f_1[b_1,b_2] = 
\sigSPS_{{\rm \textsc{nn}} \to a} \cdot \rm{f}_{\%}\,\sigma^{\rm inel}_{\rm \textsc{aa}} \cdot \langle \TAA[b_1,b_2]\rangle,\label{eq:singleAA_b}\\
\hspace{-0.5cm}\sigDPS_{\textsc{aa}[b_1,b_2]\to a b} &=& A^2 \, \sigDPS_{{\rm \textsc{nn}} \to a b}\, f_1[b_1,b_2] \times \nonumber \\
&\times&\!\!\bigg[1\!+\!\frac{2\sigmaeffdps}{A}\TAA({0})\frac{f_2[b_1,\!b_2]}{f_1[b_1,\!b_2]}+\sigmaeffdps \TAA({0})\frac{f_3[b_1,\!b_2]}{f_1[b_1,\!b_2]}\bigg],\label{eq:doubleAA_b}
\end{eqnarray}
where the latter has been obtained integrating Eq.~(\ref{eq:doubleAA1}) over $b_1~<$~$b<$~$b_2$, and 
where the three dimensionless and appropriately-normalized fractions $f_1$, $f_2$, and $f_3$ are:
\begin{eqnarray} 
f_1[b_1,b_2] &=& \frac{2\pi}{A^2} \int_{b_1}^{b_2}bdb \, \TAA({b}) = \frac{\rm{f}_{\%}\,\sigma^{\rm inel}_{\rm \textsc{aa}}}{A^2}\,\langle \TAA[b_1,b_2]\rangle,\nonumber\\
f_2[b_1,b_2] &=& \frac{2\pi}{A\, \TAA({0}) }\int_{b_1}^{b_2}bdb \int d^2b_1\, \TpA({\bf b_1}) \TpA({\bf b_1-b}) \TpA({\bf b_1-b}),\nonumber\\
f_3[b_1,b_2] &=& \frac{2\pi}{A^2\, \TAA({0}) } \int_{b_1}^{b_2}bdb\, \TAA^2({b}).\nonumber
\label{eq:f3}
\end{eqnarray}
The integrals $f_2$, and $f_3$ can be evaluated~\cite{Lokhtin:2000wm} for small enough centrality bins around a given impact parameter $b$.
The dominant $f_3$/$f_1$ contribution in Eq.~(\ref{eq:doubleAA_b}) is simply given by the ratio 
$\langle\TAA[b_1,b_2]\rangle/\TAA({0})$ which is practically insensitive (except for very
peripheral collisions) to the precise shape of the nuclear density profile. 
The second centrality-dependent DPS term, $f_2$/$f_1$, 
cannot be expressed in a simple form in terms of $\TAA({b})$, but it is of order unity for the most central
collisions, $f_2/f_1= 4/3$, and 16/15 for Gaussian and hard-sphere profiles respectively, 
and it is suppressed in comparison with the third leading term by an extra factor $\sim$2/A. 
Finally, for not very-peripheral collisions (f$_{\%}\lesssim$~0--65\%), the DPS cross section in a (thin) impact-parameter
[b$_1$,b$_2$] range can be approximated by 
\begin{eqnarray} 
\sigDPS_{\textsc{aa}\to a b}[b_1,b_2] & \approx & \sigDPS_{{\rm \textsc{nn}} \to a b}\cdot \sigmaeffdps \cdot \rm{f}_{\%} \,\sigma^{\rm inel}_{\rm \textsc{aa}} \cdot \langle \TAA[b_1,b_2]\rangle^2 \\
& = & \left(\frac{\mathpzc{m}}{2}\right)\, \sigSPS_{{\rm \textsc{nn}} \to a} \cdot \sigSPS_{{\rm \textsc{nn}} \to b} \cdot \rm{f}_{\%} \,\sigma^{\rm inel}_{\rm \textsc{aa}} \cdot \langle\TAA[b_1,b_2]\rangle^2\,.
\label{eq:doubleAA_b_final}
\end{eqnarray}
Dividing this last expression by Eq.~(\ref{eq:singleAA_b}), one finally obtains the corresponding ratio of double- to
single-parton-scattering cross sections as a function of impact parameter\footnote{Such analytical expression neglects 
the first and second terms of Eq.~(\ref{eq:doubleAA_b}). In the f$_{\%}\approx$~65--100\% centrality percentile, the second 
term would add about 20\%  more DPS cross-sections, and for very peripheral collisions (f$_{\%}\approx$~85--100\%, where 
$\langle \TAA[b_1,b_2]\rangle$ is of order or less than $1/\sigmaeffdps$) the contributions from the first term are also non-negligible.}:
\begin{eqnarray} 
(\sigDPS_{\textsc{aa}\to a b}/\sigSPS_{\textsc{aa} \to a})[b_1,b_2] \approx
\left(\frac{\mathpzc{m}}{2}\right) \, \sigSPS_{{\rm \textsc{nn}} \to b} \cdot \langle \TAA[b_1,b_2]\rangle\,.
\label{eq:doubleAA_b_ratio}
\end{eqnarray}

\subsubsection*{\mbox{DPS cross sections in AA collisions: Numerical examples}}
\label{sec:AA_DPS_ex}

Quarkonia has been historically considered a sensitive probe of the quark-gluon-plasma (QGP) formed in heavy-ion 
collisions~\cite{matsui_satz}, and thereby their production channels need to be theoretically and experimentally well 
understood in pp, pA and AA collisions~\cite{Lansberg:2008zm}.
Double-quarkonium ($\jpsi\,\jpsi$, $\Upsilon\,\Upsilon$) production is a typical channel for DPS studies in pp, 
given their large cross sections and relatively well-understood double-SPS backgrounds~\cite{Baranov:2012re,Lansberg:2014swa,Sun:2014gca}.
In Ref.~\cite{dEnterria:2013mrp}, the DPS cross section for double-$\jpsi$ production in \PbPb\ collisions 
has been estimated via Eq.~(\ref{eq:sigmaAADPS}) with $\mathpzc{m}=1$, $\sigmaeffdpsAA = 1.5 \pm 0.1$~nb, 
and prompt-$\jpsi$ SPS cross section computed at NLO via {\sc cem}~\cite{Vogt:2012vr} with the 
CT10 proton and the EPS09 nuclear PDF, and theoretical scales $\mu_{_{R}} = \mu_{_{F}} = 1.5 \,m_c$ for a $c$-quark mass 
$m_c$~=~1.27~GeV. The EPS09 nuclear modification factors result in a reduction of 20--35\% of the $\jpsi$
cross sections compared to those calculated using the free proton PDFs.
\begin{figure}[hbtp!]
\centering
\includegraphics[width=0.65\columnwidth]{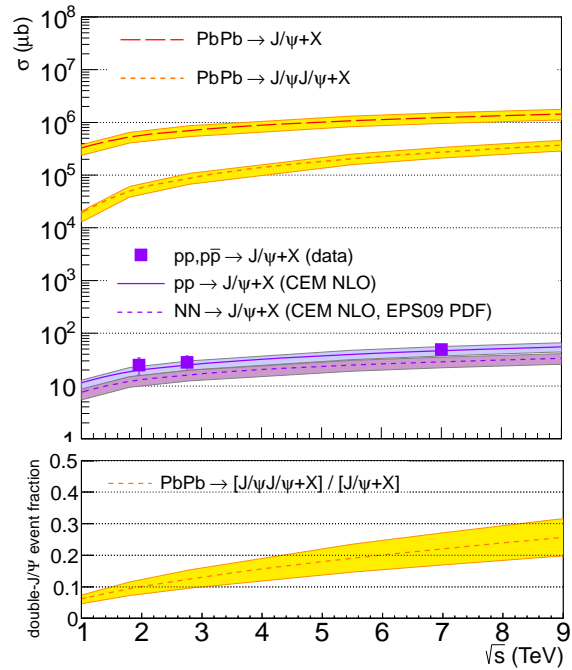}
\caption{Top: Production cross sections as a function of \cm\ energy for prompt-$\jpsi$ in \pp, NN, and \PbPb, 
and for DPS $\jpsi\,\jpsi$ in PbPb collisions. Bottom: Fraction of $\jpsi$ events in PbPb collisions with a pair 
of $\jpsi$ mesons produced, as a function of $\sqrtsnn$. Bands show the nuclear PDF and scales uncertainties 
in quadrature. [Fig. from Ref.~\cite{dEnterria:2013mrp}].}
\label{fig:sigmaDPS_vs_sqrts}
\end{figure}

Figure~\ref{fig:sigmaDPS_vs_sqrts} shows the $\sqrts$-dependence of single-$\jpsi$ in \pp, NN and PbPb collisions (top panel), 
and of double-$\jpsi$ cross sections in \PbPb, as well as the fraction of $\jpsi$ events with double-$\jpsi$ produced via DPS (bottom panel). 
Our theoretical setup with CT10 (anti)proton PDF alone agrees well with the experimental pp,\,\ppbar\ 
data~\cite{Acosta:2004yw,Abelev:2012kr,Aaij:2012ana,Abelev:2012gx,Khachatryan:2010yr,Aaij:2011jh}
extrapolated to full phase space~\cite{dEnterria:2013mrp} (squares in Fig.~\ref{fig:sigmaDPS_vs_sqrts}).
At the nominal \PbPb\ energy of 5.5~TeV, the single prompt-$\jpsi$ cross sections is $\sim$1~b, and $\sim$20\%
of such collisions are accompanied by the production of a second $\jpsi$ from a double parton interaction. 
Accounting for dilepton decays, acceptance and efficiency, which reduce the yields by a factor of 
$\sim$3$\cdot$10$^{-7}$ in the ATLAS/CMS (central) and ALICE (forward) rapidities, the visible cross section is 
$d\sigmaDPSjpsijpsi/dy|_{y=0,2} \approx$~60~nb, \ie\ about 250 double-$\jpsi$ events per unit-rapidity 
(both at central and forward $y$) are expected in the four combinations of dielectron and dimuon channels 
for a $\LumiInt$~=~1~nb$^{-1}$ integrated luminosity (assuming no net in-medium $\jpsi$ suppression or enhancement).

\begin{figure}[htpb]
\centering
\includegraphics[width=0.6\columnwidth]{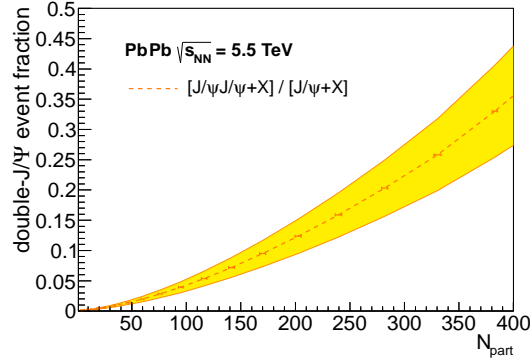}
\caption{Fraction of $\jpsi$ events in \PbPb\ collisions at 5.5~TeV where a $\jpsi$-pair is produced 
from double-parton scatterings as a function of the reaction centrality (given by N$_{\rm part}$), as per
Eq.~(\ref{eq:doubleAA_b_ratio}). The band shows the EPS09 PDF plus scale uncertainties~\cite{dEnterria:2013mrp}.
\label{fig:SPSDPS_ratio_vs_centrality}}
\end{figure}

Following Eq.~(\ref{eq:doubleAA_b_ratio}), the probability of $\jpsi\,\jpsi$ DPS production increases rapidly with 
decreasing impact parameter and $\sim$35\% of the most central \PbPb$\,\to\jpsi+X$ collisions have a second $\jpsi$ 
produced in the final state (Fig.~\ref{fig:SPSDPS_ratio_vs_centrality}). 
These results show quantitatively the large probability for double-production of $\jpsi$ mesons in 
high-energy nucleus-nucleus collisions. Thus, the observation of a $\jpsi$ pair in a given \PbPb\ event should not be 
(blindly) interpreted as \eg\ indicative of $\jpsi$ production via $\ccbar$ regeneration in the QGP~\cite{Andronic:2010dt},
since DPS constitute an important fraction of the inclusive $\jpsi$ yield, with or without final-state dense medium effects.\\


Table~\ref{tab:5} collects the DPS cross sections for the (pair) production of quarkonia ($\jpsi$, $\Upsilon$) 
and/or electroweak bosons (W, Z) in \PbPb\ collisions at the nominal LHC energy of 5.5~TeV, 
obtained via Eq.~(\ref{eq:sigmaAADPS}) with  $\sigmaeffdpsAA = 1.5$~nb.
The visible DPS yields for $\cal{L}_{\rm int}$~=~1~nb$^{-1}$ are quoted 
taking into account BR($\jpsi,\Upsilon$,W,Z)~=~6\%, 2.5\%, 11\%, 3.4\% per dilepton decay; 
plus simplified acceptance and efficiency losses: ${\cal A\times E}(\jpsi$)~$\approx$~0.01  (over 1-unit of 
rapidity at $|y|=0$, and $|y|=2$), and ${\cal A\times E}(\Upsilon;$W,Z$)\approx$~0.2; 0.5 (over $|y|<2.5$). 
All listed processes are in principle observable in the LHC heavy-ion runs, whereas rarer DPS processes
like W+Z and Z+Z have much lower visible cross sections and would require much higher luminosities
and/or \cm\ energies such as those reachable at the FCC.

\renewcommand{\arraystretch}{1.3}
\begin{table}[htbp!]
\tbl{\label{tab:5}Production cross sections at $\sqrtsnn = 5.5$~TeV for SPS quarkonia and electroweak bosons in NN collisions,
  and for DPS double-$\jpsi$, $\jpsi+\Upsilon$, $\jpsi$+W, $\jpsi$+Z,double-$\Upsilon$, $\Upsilon$+W, $\Upsilon$+Z, and same-sign WW, in PbPb.
  DPS cross sections are obtained via Eq.~(\ref{eq:sigmaAADPS}) for $\sigmaeffdpsAA = 1.5$~nb
  (uncertainties, not quoted, are of the order of 30\%), and the corresponding yields, after dilepton decays 
  and acceptance+efficiency losses (note that the $\jpsi$ yields are {\it per unit of rapidity} at mid- and forward-$y$, see text), 
  are given for the nominal 1~nb$^{-1}$ integrated luminosity.
 \vspace{0.25cm}}
{\begin{tabular}{lcccccccc}\hline
\PbPb\ (5.5 TeV)        & $\jpsi+\jpsi$ \hspace{0.5cm}& $\jpsi+\Upsilon$ \hspace{0.5cm}& $\jpsi$+W \hspace{0.5cm}& $\jpsi$+Z \hspace{0.5cm}\\\hline
$\sigSPS_{{\rm \textsc{nn}}\to a},\sigSPS_{{\rm \textsc{nn}}\to b}$  & 25~$\mu$b ($\times2$) & 25~$\mu$b, 1.7~$\mu$b & 25~$\mu$b, 30~nb & 25~$\mu$b, 20~nb \\
$\sigDPS_{\rm PbPb}$            & 210 mb & 28 mb & 500 $\mu$b & 330 $\mu$b\\
$\NDPS_{\rm PbPb}$ (1 nb$^{-1}$)\hspace{0.5cm} & $\sim$250 & $\sim$340 & $\sim$65 & $\sim$14 \\\hline
                       & $\Upsilon+\Upsilon$ & $\Upsilon+$W & $\Upsilon+$Z & ss\,WW \\\hline
$\sigSPS_{{\rm \textsc{nn}}\to a},\sigSPS_{{\rm \textsc{nn}}\to b}$  & 1.7~$\mu$b ($\times2$) & 1.7~$\mu$b, 30~nb & 1.7~$\mu$b, 20~nb & 30~nb ($\times2$) \\
$\sigDPS_{\rm PbPb}$            & 960 $\mu$b & 34 $\mu$b & 23 $\mu$b & 630 nb \\
$\NDPS_{\rm PbPb}$ (1 nb$^{-1}$)& $\sim$95 & $\sim$35 & $\sim$8 & $\sim$15 \\\hline
\end{tabular}}
\end{table}

\subsection{TPS cross sections in AA collisions}

\begin{figure}
\centering
\includegraphics[width=0.70\columnwidth,clip]{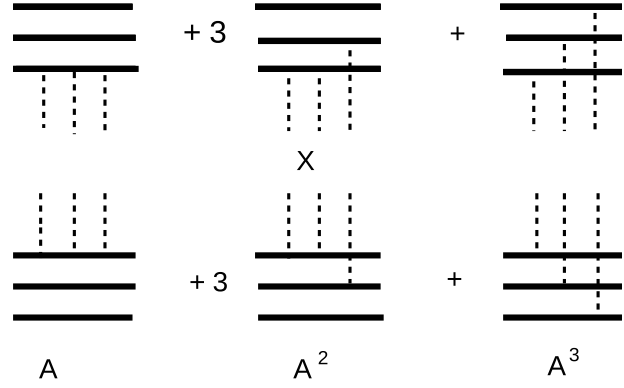}
\caption{Schematic diagrams contributing to TPS cross sections in AA collisions.}
\label{fig:TPS_AA}
\end{figure}

For completeness, we estimate here the expected scaling of TPS cross sections in nucleus-nucleus compared to proton-proton collisions. 
Following our discussion for pA in Sec.~\ref{sec:TPS_pA}, the TPS cross section in AA collisions
results from  the sum of nine terms, schematically represented in Fig.~\ref{fig:TPS_AA}, 
generated by three independent structures appearing in triple parton scatterings in \pA:
\begin{eqnarray} 
\label{eq:TPS-AA}
\sigTPS_{{\rm \textsc{aa}} \to a b c} \propto & & A  \cdot  A + 3 A \cdot  A^2 + A  \cdot A^3 \nonumber \\
& & + 3 A^2  \cdot  A + 9 A^2  \cdot  A^2 + 3 A^2  \cdot  A^3 \\
& & + A^3  \cdot  A + 3 A^3  \cdot  A^2 + A^3  \cdot  A^3.\nonumber
\end{eqnarray} 

These nine terms have different prefactors that can be expressed as a function of the nuclear 
thickness function, and the effective TPS and DPS cross sections, as done previously for the simpler pA case,
see \eg\ Eq.~(\ref{eq:triplepA}). For instance, the first $A \cdot A$ term is just the TPS cross section in NN collisions scaled by $A^2$:
\begin{eqnarray} 
\sigma^{\rm \textsc{tps}, 1}_{{\rm \textsc{aa}} \to a b c} = A^2 \cdot \sigTPS_{{\rm \textsc{nn}} \to a b c}\,,
\label{eq:AAsigmaTPS1}
\end{eqnarray} 
whereas the last $A^3 \cdot A^3$ contribution  arises from interactions of partons from three different nucleons 
in one nucleus with partons from three different nucleons in the other nucleus (\ie\ they result from triple 
{\it nucleon-nucleon} scatterings):
\begin{eqnarray} 
\sigma^{\rm \textsc{tps}, 9}_{{\rm \textsc{aa}} \to a b c} = 
\sigTPS_{{\rm \textsc{nn}} \to a b c} \cdot \sigmaefftps^2 \cdot T_{\rm 9,\textsc{aa}}\;,\;{\rm with }\;\;T_{\rm 9,\textsc{aa}} = \int d^2 r \TAA^3(\bf r)\,.
\label{eq:AAsigmaTPS9}
\end{eqnarray} 
In this latter expression, for simplicity, we omitted the $[(A-1)(A-2)/A^2]^2$ factor needed to account for
the difference between the total number of nucleon triplets and that of different nucleon triplets. The ratio 
\begin{eqnarray} 
\sigmaTPSone_{{\rm \textsc{aa}} \to a b c}/\sigmaTPSnine_{{\rm \textsc{aa}} \to a b c} \approx [2/\sigmaeffdps\, \TAA(0)]^2 
\label{eq:AAsigmaTPS1/TPS9}
\end{eqnarray} 
shows that the ``pure'' TPS contributions arising from partonic collisions within a single nucleon
(which scale as $A^2$) are negligible 
compared to triple particle production coming from three independent nucleon-nucleon collisions 
which scale as $A^6 (r_p/R_{\textsc{a}})^4\propto A^{14/3}$. In the PbPb case, the relative weights of these
two ``limiting'' TPS contributions are $1:40\,000$, to be compared with $1:200$ for the similar DPS weights.
The many other intermediate terms of Eq.~(\ref{eq:TPS-AA}) correspond to the various ``mixed'' parton-nucleon 
contributions, which can be also written in analytical form in this approach but, however, are suppressed by 
additional powers of $A$ compared to the dominant nucleon-nucleon triple scattering. 

Thus, as found in the DPS case, TPS processes in AA collisions are not so useful to derive $\sigmaeffdps$ or $\sigmaefftps$ 
and thereby study the intranucleon partonic structure as in pp or pA collisions. The estimates presented here demonstrate that double- and triple- (hard)
nucleon-nucleon scatterings represent a significant fraction of the inelastic hard AA cross section, and the standard 
Glauber MC provides a simper approach to compute their occurrence in a given heavy-ion collision.


\section{Summary}
\label{sec:summary}

Multiparton interactions are a major contributor to particle production in proton and nuclear collisions at
high center-of-mass energies. The possibility to concurrently produce multiple particles with large transverse 
momentum and/or mass in independent parton-parton scatterings in a given proton (nucleon) collision 
increases with $\sqrts$, and provides valuable information on the badly-known 3D partonic profile of hadrons, 
on the unknown energy evolution of the parton density as a function of impact parameter $b$, and on the role of 
partonic spatial, momentum, flavour, colour,... correlations in the hadronic wave functions.

We have reviewed the factorized framework that allows one to compute the cross sections for the simultaneous perturbative
production of particles in double- (DPS), triple- (TPS), and in general $n$-parton (NPS) scatterings, from the corresponding 
single-parton scattering (SPS) cross sections in proton-proton, proton-nucleus, and nucleus-nucleus collisions. 
The basic parameter of the factorized ansatz is an effective cross section parameter, $\sigmaeff$, encoding all 
unknowns about the underlying generalized $n$-parton distribution function in the proton (nucleon). In the simplest 
and most phenomenologically-useful approach, we have shown that $\sigmaeff$ bears a simple geometric interpretation 
in terms of powers of the inverse of the integral of the hadron-hadron overlap function over all impact parameters. 
Simple recursive expressions can thereby be derived to compute the NPS cross section from the $n$-th product 
of the SPS ones, normalized by $n$th$-$1 power of $\sigmaeff$.
In the case of pp collisions, a particularly simple and robust relationship between the effective DPS and TPS cross sections,
$\sigmaeffdps = (0.82\pm 0.11)\times \sigmaefftps$, has been extracted from an exhaustive analysis of typical parton 
transverse distributions of the proton, including those commonly used in Monte Carlo hadronic generators such as \pythia~8 and \herwig.

In proton-nucleus and nucleus-nucleus collisions, the parton flux is augmented by the number $A$ and $A^2$, respectively, 
of nucleons in the nucleus (nuclei). The larger nuclear transverse parton density compared to that of protons, results in enhanced 
probability for NPS processes, coming from interactions where the colliding partons belong to the same nucleon, and/or to 
two or more different nucleons. Whereas the standard SPS cross sections scale with the mass-number $A$ in pA relative to pp collisions, 
we have found that the DPS and TPS cross sections are further enhanced by factors of order $(A+(1/\pi)\,A^{4/3})$ and 
$(A+(2/\pi)\,A^{4/3}+(1/\pi^2)\,A^{5/3})$ respectively. 
In the case of pPb collisions, this implies enhancement factors of $\sim$600 (for DPS) and of $\sim1900$ (for TPS) with respect to the corresponding SPS 
cross sections in pp collisions. The relative roles of intra- and inter-nucleon parton contributions to DPS and TPS
cross sections in pA collisions have been also derived. In pPb, 1/3 of the DPS yields come from partonic interactions 
within just one nucleon of the Pb nucleus, whereas 2/3 involve scatterings from partons of two Pb nucleons; whereas for 
the TPS yields, 10\% of them come from partonic interactions within one nucleon, 50\% involve scatterings within 
two nucleons, and 40\% come from partonic interactions in three different Pb nucleons. In proton-nucleus collisions,
one can thereby exploit the large expected DPS and TPS signals over the SPS backgrounds to study double- and triple- parton 
scatterings in detail and, in particular, to extract the value of the key $\sigmaeffdps$ parameter independently of measurements in pp 
collisions, given that the corresponding NPS yields in pA depend on the comparatively better-known nuclear transverse density profile.

For heavy ions, the $A^2$-scaling of proton-proton SPS cross sections becomes $\propto(A^2+(2/\pi)\,A^{7/3}+1/(2\pi)\,A^{10/3})$ 
for DPS cross sections, and includes much larger powers of $A$ (up to $A^{14/3}$) for TPS processes. In the PbPb case, these 
translate into many orders-of-magnitude enhancements 
(\eg\ the DPS cross sections are $\sim10^7$ larger than the corresponding SPS pp ones). In addition, the MPI probability is significantly enhanced for 
increasingly central collisions: the impact-parameter dependence of DPS cross sections is basically proportional to the AA nuclear 
overlap function at a given $b$. The huge DPS and TPS cross sections expected in AA collisions are, however, clearly dominated 
by scatterings among partons of {\it different} nucleons, rather than by partons belonging to the same proton or neutron.
For nuclei with mass numbers $A = 40$--240, the relative weights of the three components contributing to DPS scattering in 
AA collisions are $1:2.3:23$ (for $A = 40$) and $1:4:200$ (for $A = 208$). Namely, only 13\% (for $^{40}$Ca+$^{40}$Ca) or 
2.5\% (for $^{208}$Pb+$^{208}$Pb) of the DPS yields in AA collisions come from diagrams involving partons from one single nucleon. 
Clearly, the ``pure'' DPS contributions involving partonic collisions within a nucleon are much smaller than those 
issuing from two independent {\it nucleon-nucleon} collisions. In the TPS case, the relative weights of the
two extreme contributions (three parton collisions within two single nucleons versus those from three different nucleon-nucleon collisions) 
are $1:40\,000$ for PbPb. The NPS cross sections in AA are practically unaffected by the value of $\sigmaeffdps$
and, although DPS and TPS processes account for a significant fraction of the inelastic hard AA cross section, 
they are not as useful as those in pp or pA collisions to 
study the partonic structure of the proton (nucleon).

Numerical examples for the cross sections and visible yields expected for the concurrent DPS and TPS production of heavy-quarks, 
quarkonia, and/or gauge bosons in proton and nuclear collisions at LHC, FCC, and at ultra-high cosmic-ray energies 
have been provided. The obtained DPS and TPS cross sections are based on perturbative QCD predictions for the 
corresponding single inclusive processes at NLO or NNLO accuracy including, 
when needed, nuclear modifications of the corresponding  parton densities.
Processes such as double-$\jpsi$, $\jpsi\,\Upsilon$, $\jpsi$\,W, $\jpsi\,$Z, double-$\Upsilon$, $\Upsilon\,$W, $\Upsilon\,$Z, and same-sign W\,W
production have large cross sections and visible event rates for the nominal LHC and FCC luminosities. The study of
such processes in proton-nucleus collisions provides an independent means to extract the effective $\sigmaeffdps$ parameter 
characterising the transverse parton distribution in the nucleon. In addition, we have shown that double-$\jpsi$ and double-$\Upsilon$ final states
have to be explicitly taken into account in any event-by-event analysis of quarkonia production in heavy-ion collisions. 
The TPS processes, although not observed so far, have visible cross sections for charm and bottom in pp and pA 
collisions at LHC and FCC energies. At the highest \cm\ energies reached in collisions of cosmic rays with the nuclei
in the upper atmosphere, the TPS cross section for triple charm-pair production equals the total p-Air inelastic cross section, 
indicating that {\it all} such collisions produce at least three $\ccbar$-pairs in multiple partonic interactions. The results 
presented here emphasize the importance of having a good understanding of the NPS dynamics in hadronic collisions at current and future colliders, 
both as genuine probes of QCD phenomena and as backgrounds for searches of new physics in rare final-states with multiple heavy-particles,
and their relevance in our comprehension of ultrarelativistic cosmic-ray interactions with the atmosphere.

%

\bibliographystyle{ws-rv-van}


\end{document}